\documentclass[10pt]{article}

\usepackage{graphicx}
\usepackage{amsfonts}
\usepackage{amsmath}
\usepackage{amssymb}
\usepackage{amsbsy}
\usepackage{subfigure}
\usepackage{natbib}
\usepackage{color}

\newcommand{\EQ}{\begin{equation}}
\newcommand{\EN}{\end{equation}}
\newcommand{\EQA}{\begin{eqnarray}}
\newcommand{\ENA}{\end{eqnarray}}

\newcommand{\BB}{{\bf {B}}}
\newcommand{\JJ}{{\bf {J}}}
\newcommand{\EE}{{\bf {E}}}
\newcommand{\vv}{{\bf {v}}}
\newcommand{\XX}{{\bf {X}}}

\usepackage{color}

\begin{document}

\title{Generalised models for torsional spine and fan magnetic reconnection}

\author{D.~I.~Pontin{$^*$}, A.~K.~Al-Hachami\footnote{Division of Mathematics, University of Dundee, Nethergate, Dundee, DD1 4HN, U.K.} ~\& K. Galsgaard\footnote{Niels Bohr Institute, Blegdamsvej 17, Dk-2100 Copenhagen {\O}, Denmark}}
\date{}

\maketitle

\begin{abstract}
\noindent
{{\it Context:} Three-dimensional (3D) null points are present in abundance in the solar corona, and the same is likely to be true in other astrophysical environments. Recent results from solar observations and from simulations suggest that reconnection at such 3D nulls may play an important role in the coronal dynamics.}\\
{{\it Aims:}  The properties of the torsional spine and torsional fan modes of magnetic reconnection at 3D nulls are investigated. New analytical models are developed, which for the first time include a current layer that is spatially localised around the null, extending along either the spine or the fan of the null. The principal aim is to investigate the effect of varying the degree of asymmetry of the null point magnetic field on the resulting reconnection process -- where previous studies always considered a non-generic radially symmetric null.}\\
{{\it Methods:} Analytical solutions are derived for the steady kinematic equations, and are compared with the results of numerical simulations in which the full set of resistive MHD equations is solved.}\\
{{\it Results:} The geometry of the current layers within which torsional spine and torsional fan reconnection occur is strongly dependent on the symmetry of the magnetic field. Torsional spine reconnection occurs in a narrow tube around the spine, with elliptical cross-section when the fan eigenvalues are different. The eccentricity of the ellipse increases as the degree of asymmetry increases, with the short axis of the ellipse being along the strong field direction. The spatiotemporal peak current, and the peak reconnection rate attained, are found not to depend strongly on the degree of asymmetry. For torsional fan reconnection, the reconnection occurs in a planar disk in the fan surface, which is again elliptical when the symmetry of the magnetic field is broken. The short axis of the ellipse is along the weak field direction, with the current being peaked in these weak field regions. The peak current and peak reconnection rate in this case are clearly dependent on the asymmetry, with the peak current increasing but the reconnection rate decreasing as the degree of asymmetry is increased.}
\end{abstract}

\section{Introduction}

There has been significant progress in recent years towards understanding the properties of magnetic reconnection in three dimensions (3D). In particular, it is now appreciated that the intense current layers necessary for reconnection in astrophysical plasmas may form at a number of different characteristic structures of the magnetic field. One such structure is a 3D magnetic null point. Such null points have been demonstrated to be present in abundance in the solar corona \citep[e.g.][]{longcope2009}, and the same is highly likely to be true in other astrophysical environments such as the coronae of other stars and of accretion disks. Recent results from observations and simulations suggest that reconnection at these nulls may play an important role in the coronal dynamics \citep[e.g.][]{luoni2007,lynch2008,pariat2009,masson2009}. Note also that the importance of reconnection at 3D nulls is not restircted to astrophysical plasmas, but plays a role both in the Earth's magnetosphere \citep[e.g.][]{xiao2006} and some laboratory plasmas \citep[e.g.][]{gray2010}.

Recent studies have revealed a number of characteristic modes of reconnection that may occur at such 3D nulls. 
These have recently been categorised by \cite{priest2009} into `torsional spine reconnection', `torsional fan reconnection' and `spine-fan reconnection'. Torsional spine and torsional fan reconnection occur when an equilibrium magnetic null point field is disturbed by a rotational perturbation (the rotation being in a plane perpendicular to the spine). Spine-fan reconnection occurs when a shear perturbation is applied that disturbs the locations of the spine and fan -- this leads to a localised current layer forming around the null and flux transport across the fan separatrix surface.

Close to the null the magnetic field may be written 
\begin{equation}
\BB = \nabla\BB\cdot {\bf x}
\end{equation}
and the eigenvalues and eigenvectors of the matrix $\nabla\BB$ determine the locations of the spine and fan of the null \citep[see e.g.][]{parnell1996}.
Previous studies of the generation of current layers at 3D nulls due to rotational motions have considered only the case where the background equilibrium null point has a rotational symmetry, i.e. in which the two eigenvalues associated with the fan are equal. 
During the resulting evolution in which torsional spine and torsional fan reconnection takes place, the respective current layers, plasma flows, etc., have exhibited azimuthal symmetry due to the azimuthally symmetric background magnetic fields and perturbations.We have previously investigated the effect of varying the symmetry of the background null point field on the spine-fan reconnection mode, and shown that while the topological properties of the reconnection are unchanged, the dimensions and intensity of the current layer, and the reconnection rate, are strongly dependent on the degree of asymmetry \citep{alhachami2010}, as well as the relative angle between the shear driving and the strong/weak field directions \citep{galsgaard2011b}.
In this paper we generalise existing models for torsional spine and torsional fan null point reconnection as follows. We begin in each case by introducing a new kinematic analytical  solution for the corresponding reconnection mode in which a localised current layer is present at the null. We then go on to consider the effect of varying the symmetry of the background magnetic field, by varying the ratio of the fan eigenvalues of the null. Lastly we perform numerical simulations of the full system of MHD equations in which null point configurations with varying degrees of symmetry are subjected to rotational disturbances, to complement the analytical models.

The paper is organised as follows. In Sect.~\ref{previous.sec} we review previous modelling of torsional spine and torsional fan  reconnection. In Sect.~\ref{tspine.sec}  we introduce our new analytical and numerical models for torsional spine reconnection, and discuss their implications. We go on in Sect.~\ref{tfan.sec}  to do the same for torsional fan reconnection. Finally in Sect.~\ref{conc.sec} we present our conclusions.

\section{Existing models for torsional spine and fan reconnection}\label{previous.sec}
Both torsional spine and torsional fan reconnection involve the formation of a current layer in which the current vector is directed parallel to the spine line at the null. It was first shown by \cite{pontin2004} that the corresponding reconnection takes the form of a rotational slippage of magnetic flux threading the non-ideal region. (This is in contrast to the case where the current vector is parallel to the fan, in which case magnetic flux is reconnected across the spine and fan \citep{pontinhornig2005}.) The magnetic flux undergoes this rotational slippage in response to rotational flows in the ideal region in which the rate of flux transport in the azimuthal direction is different for field lines entering the non-ideal region than it is for field lines exiting the non-ideal region. The original model of \cite{pontin2004} is based on the magnetic field
$$
\BB=B_0 [r,jr/2,-2z]
$$
in cylindrical polar coordinates. This results in a spatially uniform current parallel to the spine ($z$-axis). In order to obtain a 3D-localised non-ideal region (relevant in astrophysical plasmas which are approximately ideal almost everywhere), it was therefore necessary to impose a localised resistivity profile. 

In the following sections we introduce two new analytical kinematic solutions. In these solutions we have been able, for the first time, to include a current density which is fully spatially localised with its peak intensity focussed at the spine or fan. This current localisation allows a spatially uniform resistivity to be used, adding a degree of physical plausibility to the models, since in practice in an astrophysical plasma a localised non-ideal region is associated with a localised current layer. In each of our solutions, the structure chosen for the magnetic field and the resulting current layer is based on the results of resistive MHD simulations in which the dynamic formation of these current layers has been observed. 
It is worth noting that additional kinematic solutions with a spatially varying current density have been presented by  \cite{wyper2010}. In their solutions the current is spatially localised, but in contrast to our new solutions is  focussed away from the null point.

Resistive MHD simulations have demonstrated that the form of the current layer is different depending on whether the rotational perturbation primarily disturbs the fan field lines or field lines around the spine. The perturbation behaves essentially as an Alfv{\' e}n wave, travelling along the magnetic field lines, which due to the hyperbolic structure of the field around the null leads to the perturbation accumulating either in the vicinity of the spine or the fan. When the fan field lines are subjected to a rotational disturbance, torsional spine reconnection takes place in a tubular current structure that forms at the spine \citep{rickard1996,pontingalsgaard2007}. Within this tube, the magnetic field lines spiral around the spine line. The radius of the tube decreases, and the current intensifies, until the twisting of the field being driven by the perturbation is balanced by rotational slippage facilitated by resistive diffusion. The reconnection rate, defined in 3D as the maximal value of 
\EQ \label{recrate.eq}
\Psi=\int \EE\cdot\BB/|\BB|\, dl
\EN
along any field line threading the non-ideal region, then quantifies this rotational slippage. 

When field lines in the vicinity of the spine line are disturbed, a current layer forms on the fan surface leading to torsional fan reconnection \citep{rickard1996,galsgaardpriest2003}. Again field lines spiral within the current layer, whose magnitude intensifies as the twisting of the field is concentrated in an increasingly narrow sheet over the fan surface. Once the sheet becomes sufficiently thin resistive diffusion dissipates the twist leading once again to a rotational slippage of field lines.

\section{Torsional spine reconnection}\label{tspine.sec}
\subsection{Torsional spine reconnection: kinematic model}\label{tspine.kin}
In this section we describe a kinematic model for torsional spine reconnection within a current tube that is localised to the spine of the null point. The form of the magnetic field and resulting current structure is chosen to match behaviour observed in the numerical simulations described by \cite{pontingalsgaard2007}. The method of solution is the same as that described by \cite{hornig2003} and \cite{pontin2004}. Specifically, we solve the steady-state, kinematic, resistive MHD equations 
\EQ \label{kineq.eq}
\begin{array}{ccc}
\EE+\vv\times\BB=\eta\JJ &~~~~& \nabla\times\BB=\mu_0\JJ \\
\nabla\times\EE={\bf 0} && \nabla\cdot\BB=0.
\end{array}
\EN
We specify the form of the magnetic field $\BB$ and solve for the electric field ($\EE$) and plasma flow perpendicular to $\BB$ ($\vv_\perp$) via
\EQ
\label{kinsol}
\Phi=\int \eta \JJ\cdot\BB \, ds, \qquad \EE=-\nabla\Phi, \qquad \vv_{\perp} = \frac{(\EE-\eta\JJ)\times\BB}{B^2},
\EN
(the component of $\vv$ parallel to $\BB$ being arbitrary). Note that $s$ is a parameter along magnetic field lines, i.e.~the integration of $\Phi$ is performed along these field lines. The resistivity $\eta=\eta_0$ is taken to be uniform.

\begin{figure} \centering
 \includegraphics[width=0.35\textwidth]{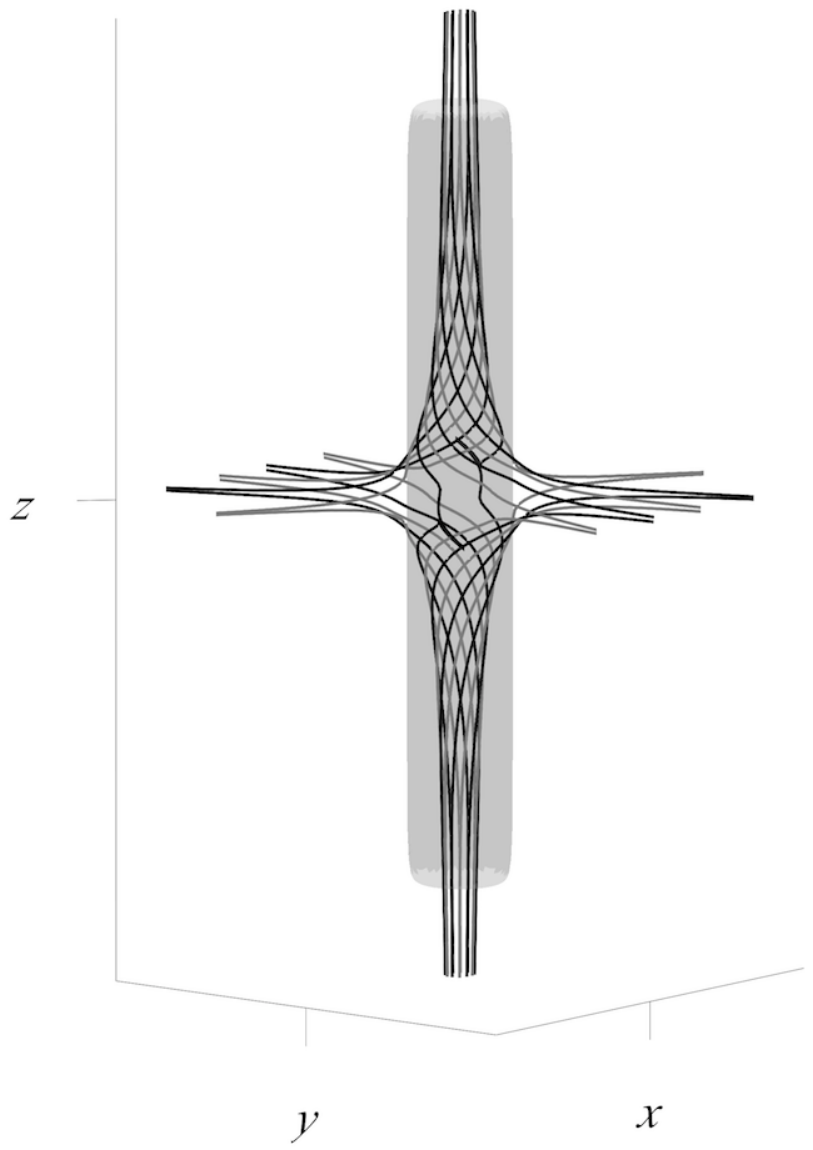}
 \includegraphics[width=0.33\textwidth]{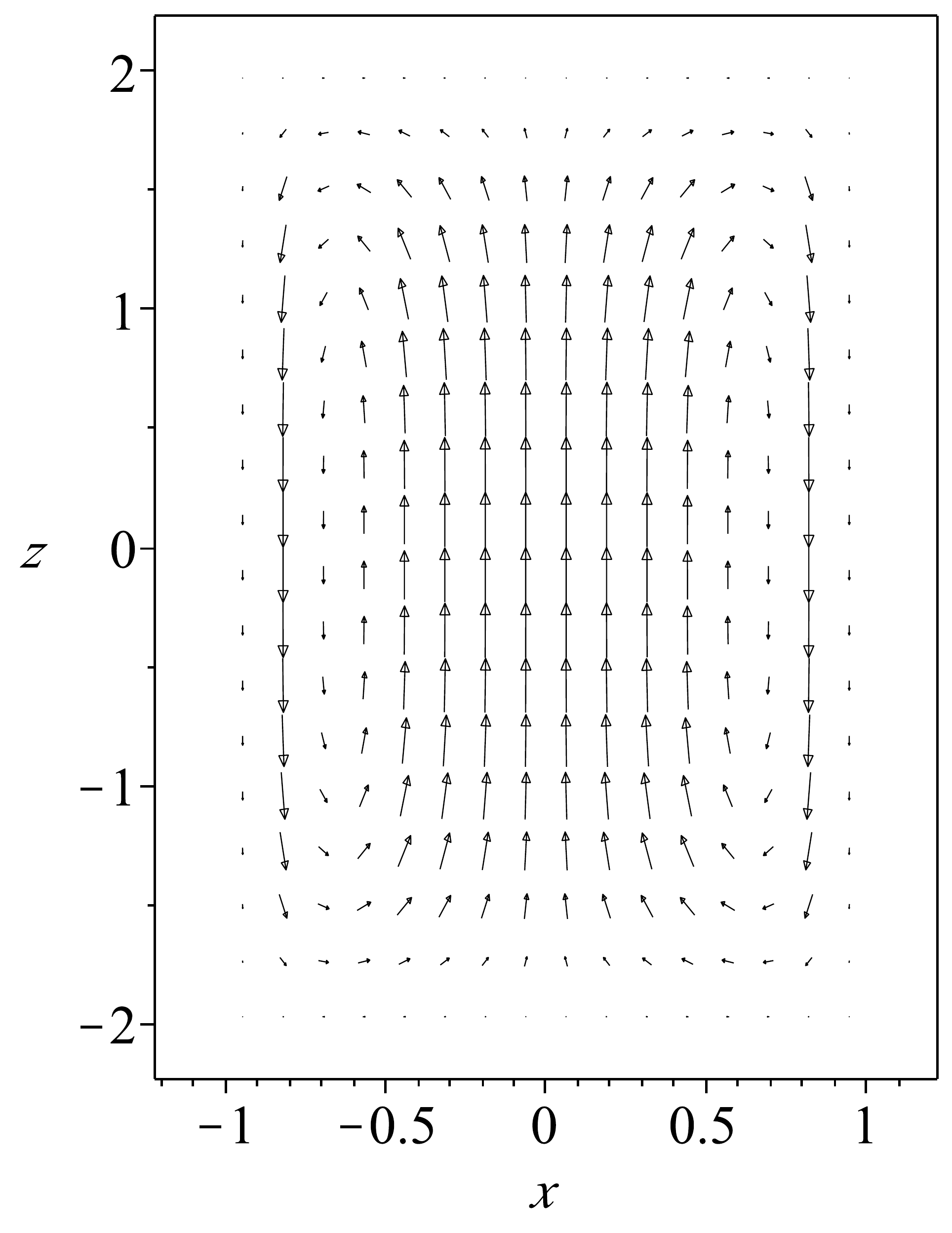} 
\caption[]{\label{kin_tspine.fig} {\it Left}: Magnetic field lines for the torsional spine model defined by Eqs.~(\ref{totb.eq},\ref{anis_null.eq},\ref{anis_spineb.eq}) for $a=1,b=5,j=3,p=1.5,q=1$. The shaded surface shows a current isosurface. {\it Right}: current vectors in the $y=0$ plane for $a=1,b=2,q=1$.}
\end{figure}

We consider first the rotationally symmetric case. We set
\EQ
\BB=\frac{B_0}{L_0}(\BB_P+\BB_J) \label{totb.eq}
\EN
where
\EQA
\BB_P \!&=&\! [r, 0, -2z] \label{potb.eq}\\
\BB_J \!&=&\! \left\{ 
\begin{array}{cc}
\left[ 0, jr \left(1-\left(\frac{r}{a}\right)^{2m}\right)^{2\mu}\left(1-\left(\frac{z}{b}\right)^{2n}\right)^{2\nu},0 \right] &~~~ 
\begin{array}{c}
r\leq a \\ ~\&~ |z|\leq b
\end{array} \\
 \rule{0pt}{3ex}
\left[0,0,0 \right] & ~~~ {\rm otherwise}
\end{array}
\right. \label{spineb.eq}
\ENA
in cylindrical polar coordinates $(r,\theta,z)$ where $j, a, b$ are positive real numbers and $n,m,\mu,\nu$ are positive integers. $\BB_P$ defines the potential `background' null point component, and $\BB_J$ defines the non-potential component of $\BB$ associated with the tubular current structure, which extends to radius $r=a$ and to $z=\pm b$. In accordance with previous simulation results we assume an extended tube of current aligned to the spine so that  $b\gg a$. For $m=n=\mu=\nu=1$, $\BB$ and $\JJ$ are continuous and differentiable. However, in order that all physical quantities in the final solution are continuous and differentiable we find that it is necessary to choose  $m=3, \mu=n=2$ and $\nu=1$. The current density is represented by the plot in Fig.~\ref{kin_tspine.fig}.

\begin{figure} \centering
\includegraphics[width=0.34\textwidth]{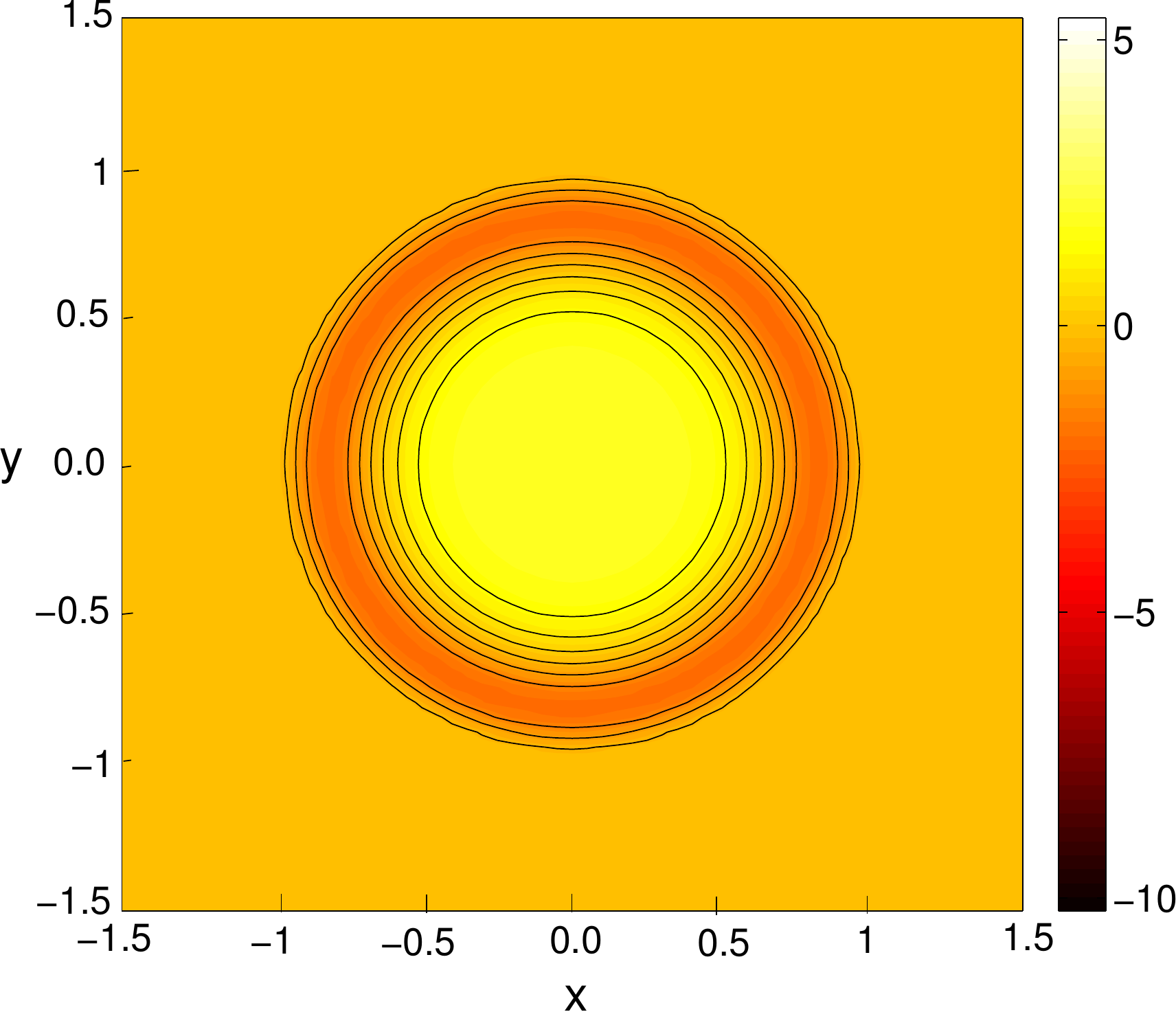}
\includegraphics[width=0.34\textwidth]{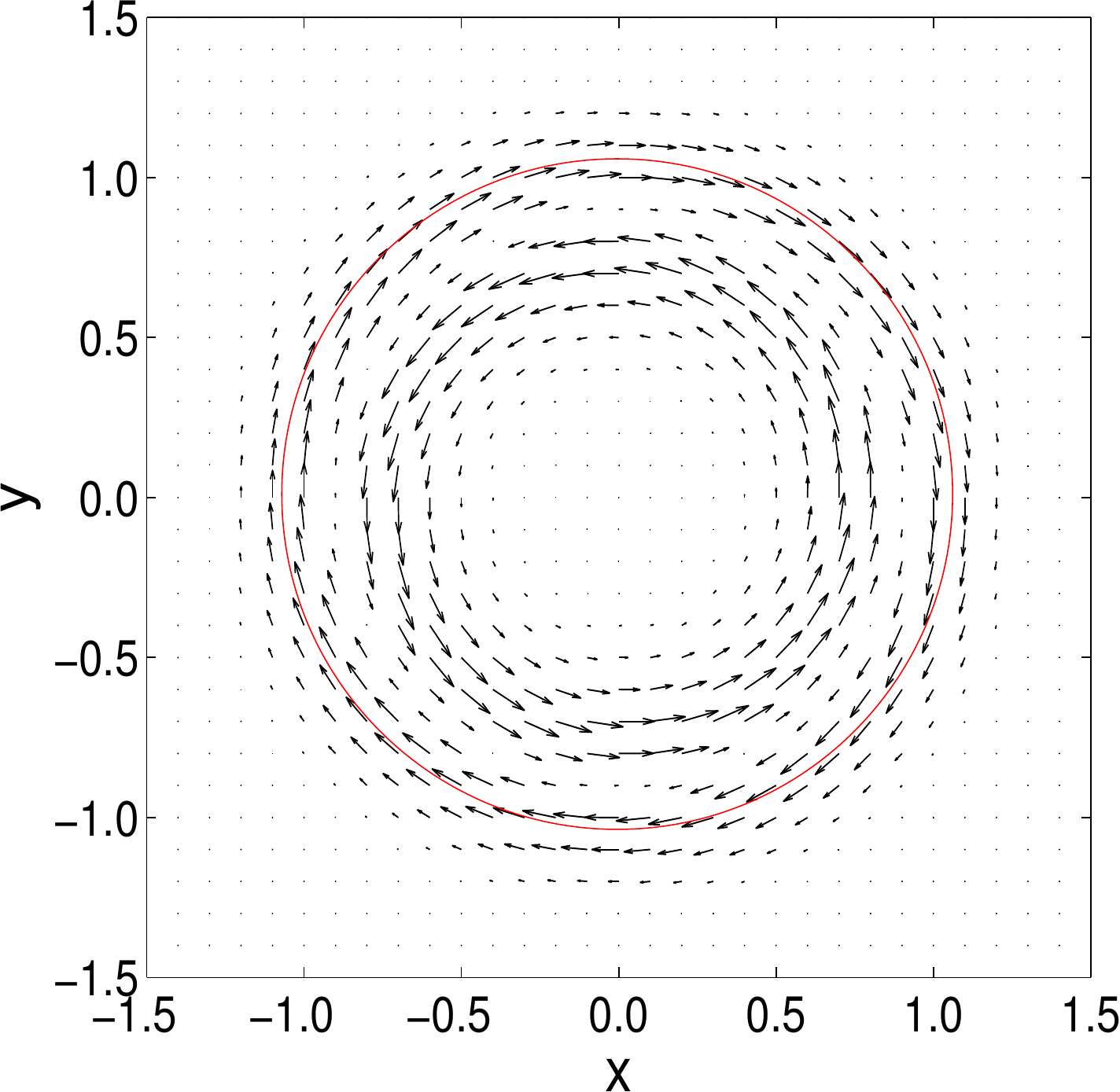}\\
\includegraphics[width=0.34\textwidth]{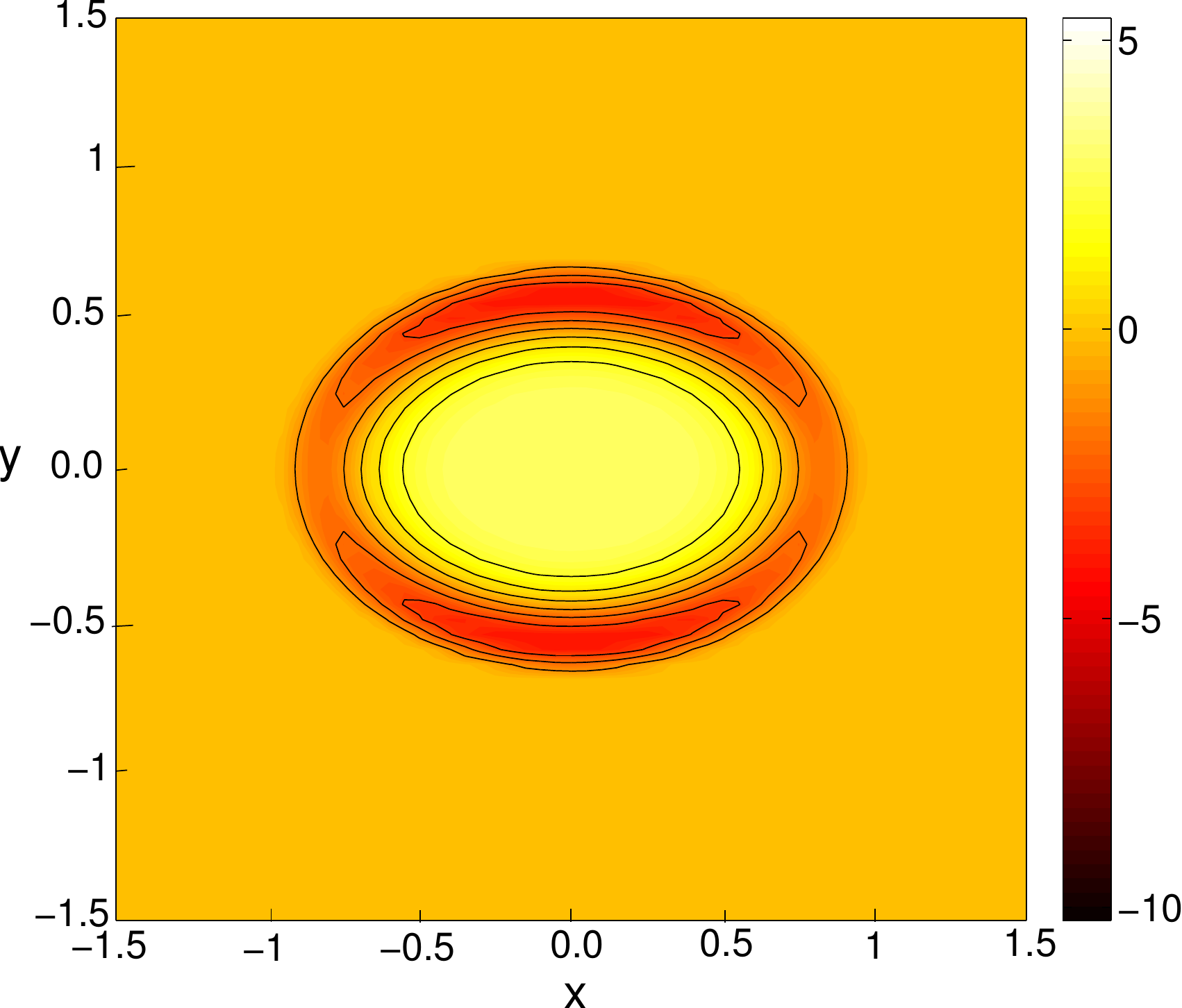}
\includegraphics[width=0.34\textwidth]{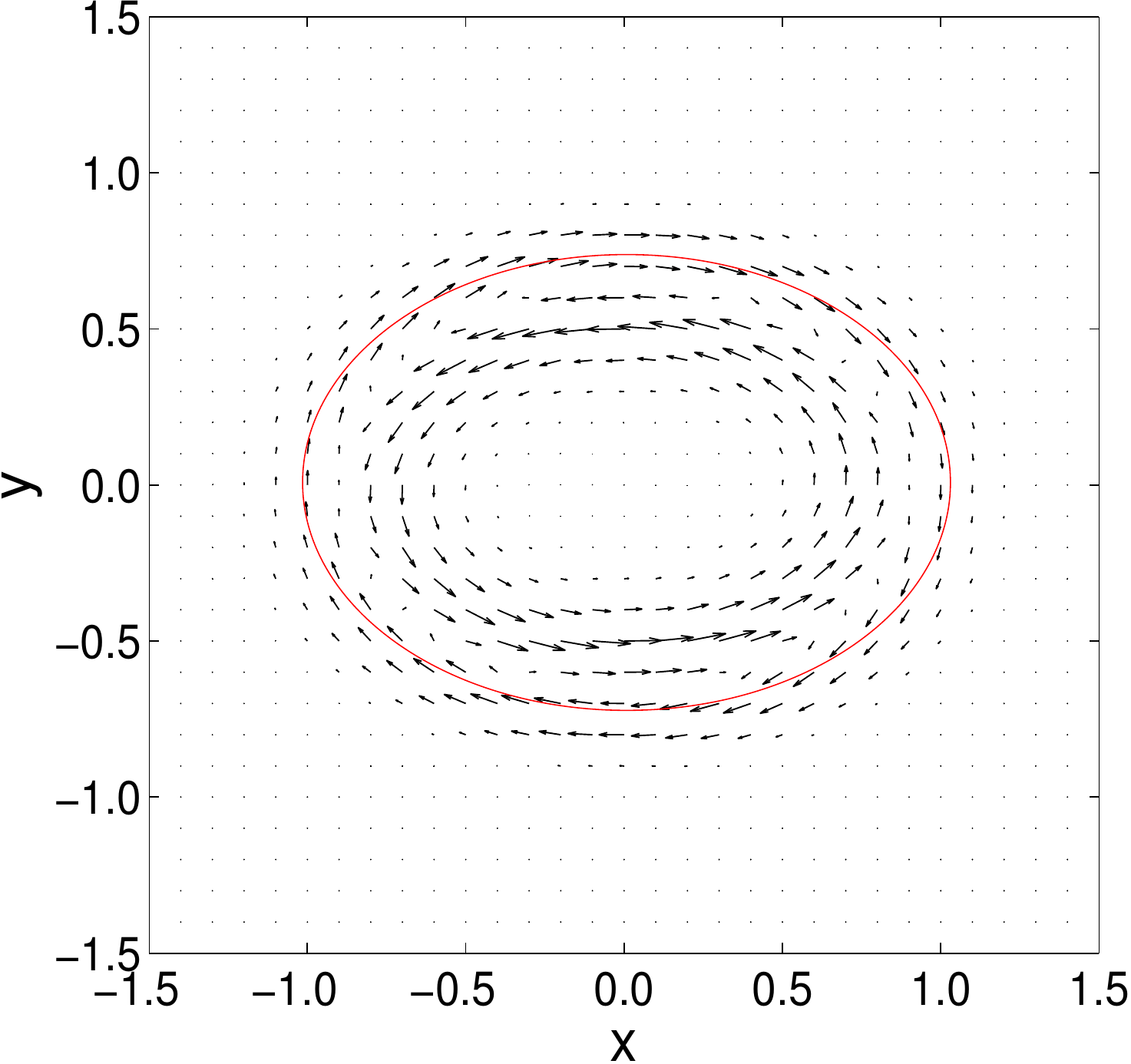}\\
\includegraphics[width=0.345\textwidth]{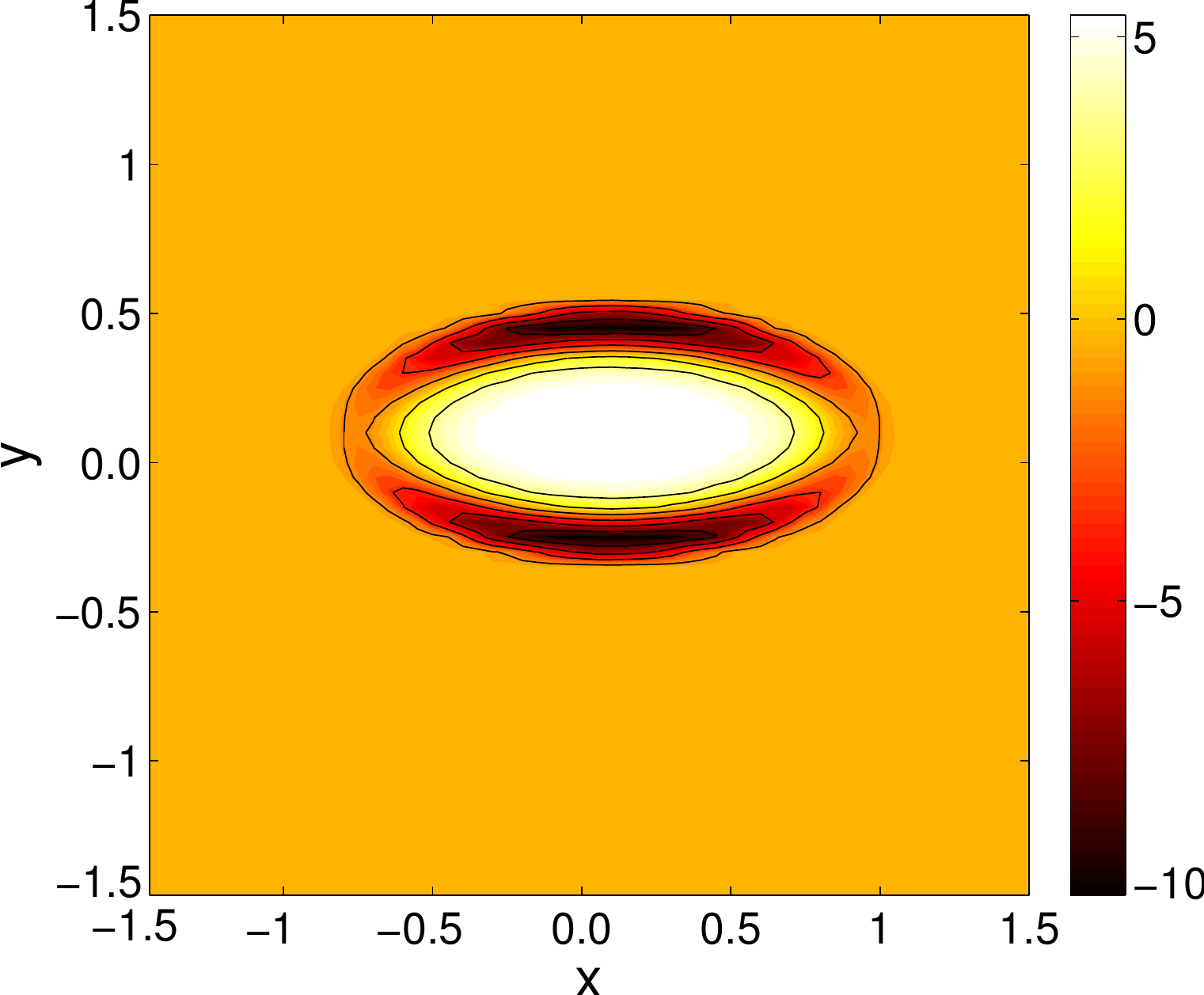}
\includegraphics[width=0.34\textwidth]{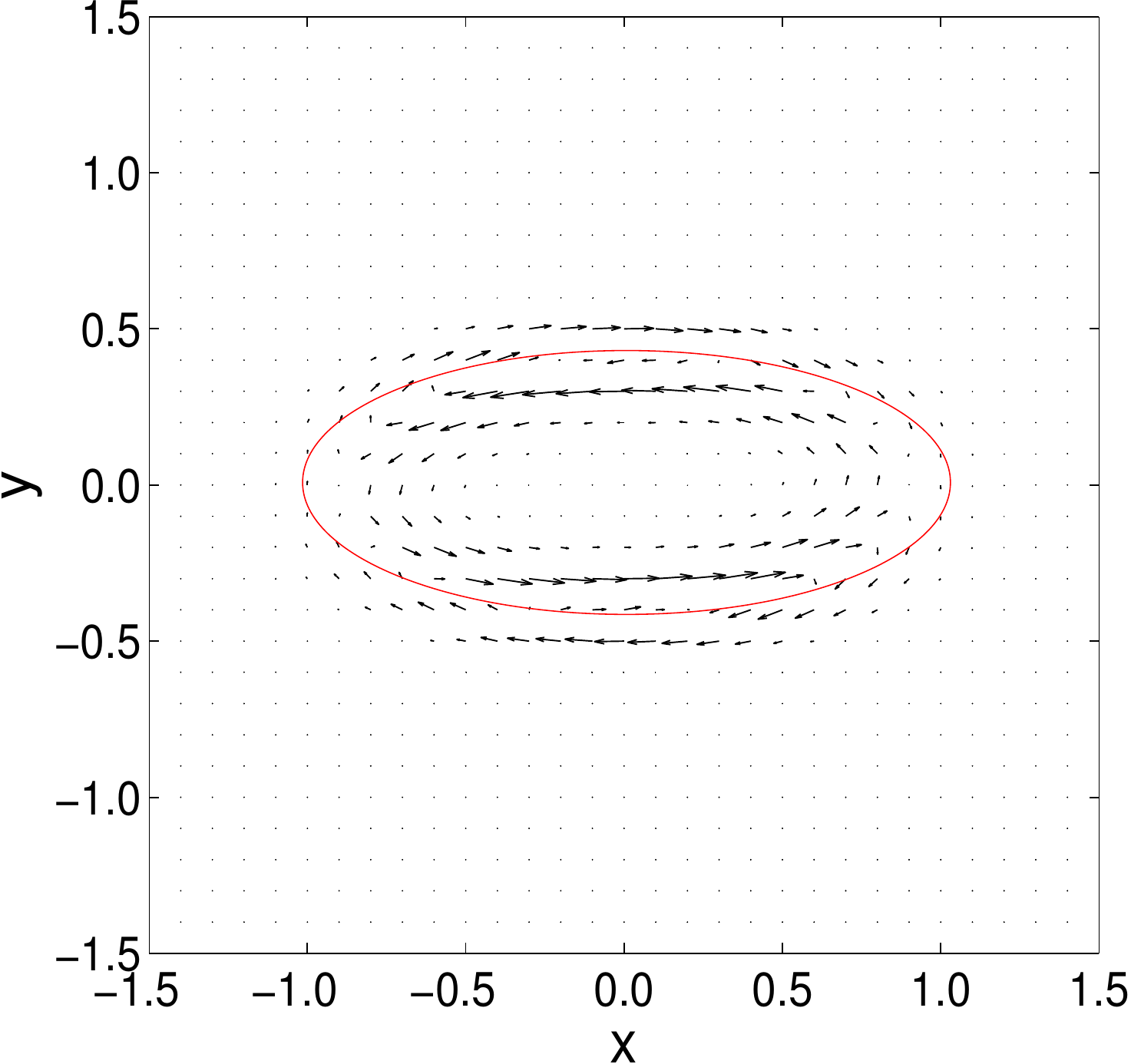}
\caption[]{\label{res_tspine.fig} Current density (left) and plasma flow vectors (right) for the torsional spine kinematic solution. Plotted in the plane $z=2$ for $B_0=L_0=j=a=1, b=4$ and $p=q=1$ (above), $p=q=2$ (middle) $p=q=5$ (below). The grey curve (red online) in the right-hand plots marks the boundary of the non-ideal region.}
\end{figure}

Parametric equations for the magnetic field lines associated with Eqs.~(\ref{totb.eq}-\ref{spineb.eq}) can be found in a straightforward way by solving $\partial\XX/\partial s=\BB(\XX(s))$. Since the integrand in the first equation in (\ref{kinsol}) is independent of $\theta$ we require only $r=r_0 \exp(B_0 s/L_0)$ and $z=z_0 \exp(-2B_0 s/L_0)$ (it is straightforward to also obtain an expression for $\theta(r_0,\theta_0,z_0,s)$). Solving Eqs.~(\ref{kinsol}) with $s=0$ at $z=\pm z_0$ we find that the qualitative properties of the electric field and flow field are similar to those discovered by \cite{pontin2004}. In particular, rotational plasma flows are still present around the spine axis of the null. Using the freedom of arbitrary flow parallel to $\BB$ in the model we can  for illustrative purposes choose to add a component $\vv_\|$ such that $v_z=0$. We then see a purely azimuthal flow.
 Due to the return current region close to $r=a$ (see Fig.~\ref{kin_tspine.fig}) the sense of rotation of the plasma changes at an intermediate radius, as shown in the top-right frame of Fig.~\ref{res_tspine.fig}. 

We now  investigate how the properties of the solution vary when the rotational symmetry of the above system is broken. When the rotational symmetry is lost it is no longer possible to find closed-form expressions for the field line mapping. We therefore numerically integrate $\BB$ to find field lines and solve Eqs.~(\ref{kinsol}) on a rectangular grid. We may break the symmetry either in the potential component $\BB_P$ defining the magnetic null or in the component $\BB_J$ defining the current tube.  Our new potential component of the magnetic field is 
given by 
\EQ \label{anis_null.eq}
\BB_P=\left[ \frac{2}{p+1} x, \frac{2p}{p+1}y, -2z\right]
\EN
in Cartesian coordinates where $p>0$ is a parameter. As $p$ varies the magnetic field along the spine direction is fixed while the ratio between the fan eigenvalues (associated with the eigenvectors along the ${\hat {\bf x}}$ and ${\hat {\bf y}}$ directions) varies (see Fig.~\ref{kin_tspine.fig}). We choose to break the symmetry in $\BB_J$ by converting to Cartesian coordinates and setting 
\EQ
\BB_J = 
 \left\{ 
\begin{array}{cc}
j \left(1-\left(\frac{R}{a}\right)^{6}\right)^{4}\left(1-\left(\frac{z}{b}\right)^4\right)^{2}
\left[ -qy, x ,0 \right] &~~~ R\leq a ~\&~ |z|\leq b \\
\rule{0pt}{3ex}
\left[0,0,0 \right] & ~~~ {\rm otherwise}
\end{array}
\right. \label{anis_spineb.eq}
\EN
where $R^2=x^2+qy^2$ (note that this reduces to expression (\ref{spineb.eq}) when $q=1$). 
This has the effect of distorting the current into a cylinder with elliptical cross-section, with major and minor axes along the $x$- and $y$-axes, extending to $x=\pm a$, $y=\pm a/\!\sqrt{q}$ (see the images on the left in Fig.~\ref{res_tspine.fig}). 
Pre-empting the results of the following section, we present here results for $p=q$, such that as $p$ increases the current tube narrows along the direction associated with the large fan eigenvalue, i.e.~the strong field direction in the fan. We set $B_0=L_0=\eta_0=j=1$, $a=1$, $b=4$, and solve Eqs.~(\ref{kinsol}) on a rectangular grid with 81 gridpoints in each direction covering the volume $-2<x,y<2, 0<z<4$ with the solution being symmetric about $z=0$. We restrict our attention to the range $p \geq 1$, which simply selects the ${\hat {\bf y}}$ direction as the strong field direction in the fan.

\begin{figure} \centering
\includegraphics[width=0.35\textwidth]{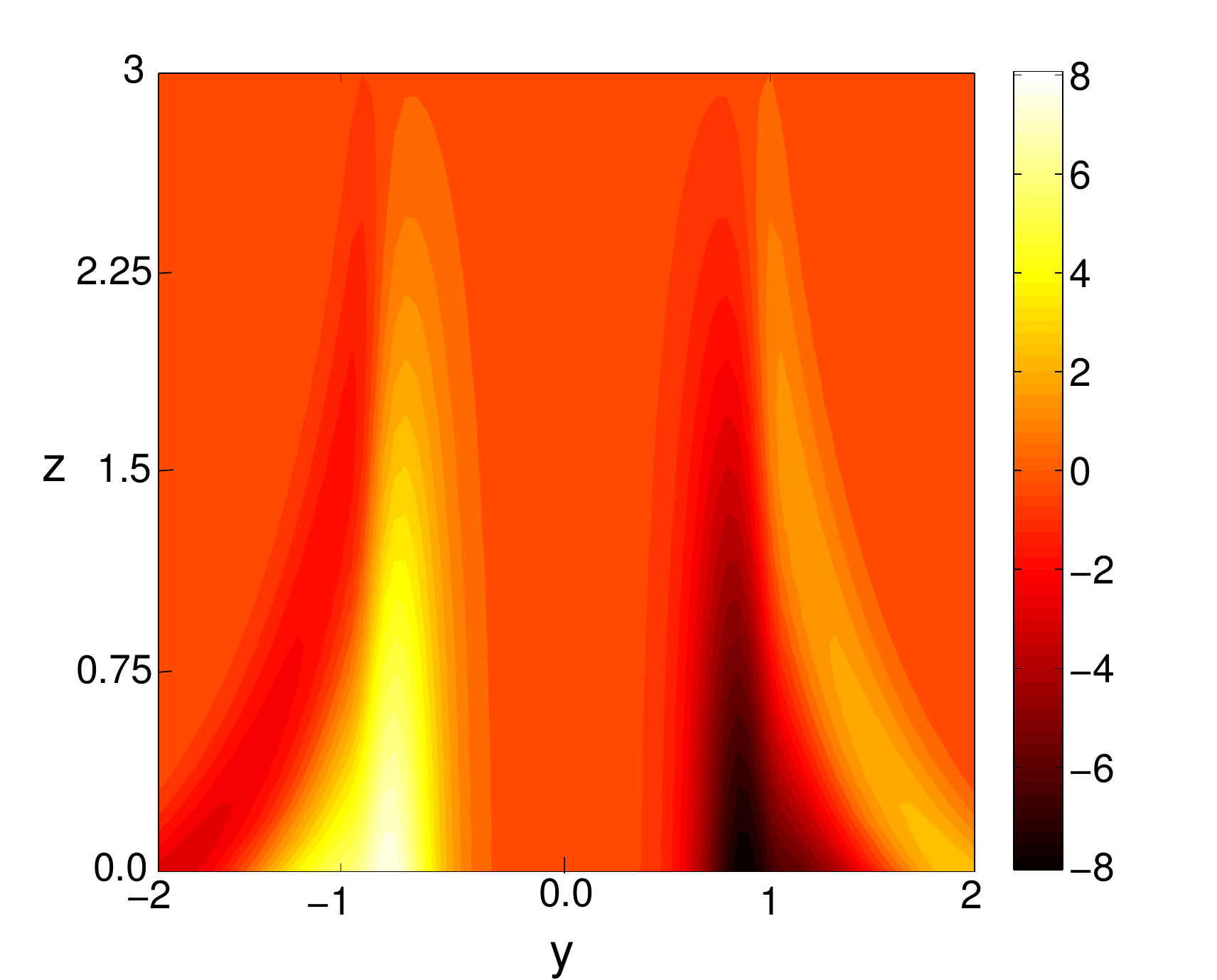}
\caption[]{\label{res_tspine_vdepz.fig} Intensity map of $v_x$ in the $x=0$ plane for the kinematic torsional spine solution with $p=q=1$.}
\end{figure}

\begin{figure} \centering
\includegraphics[width=0.35\textwidth]{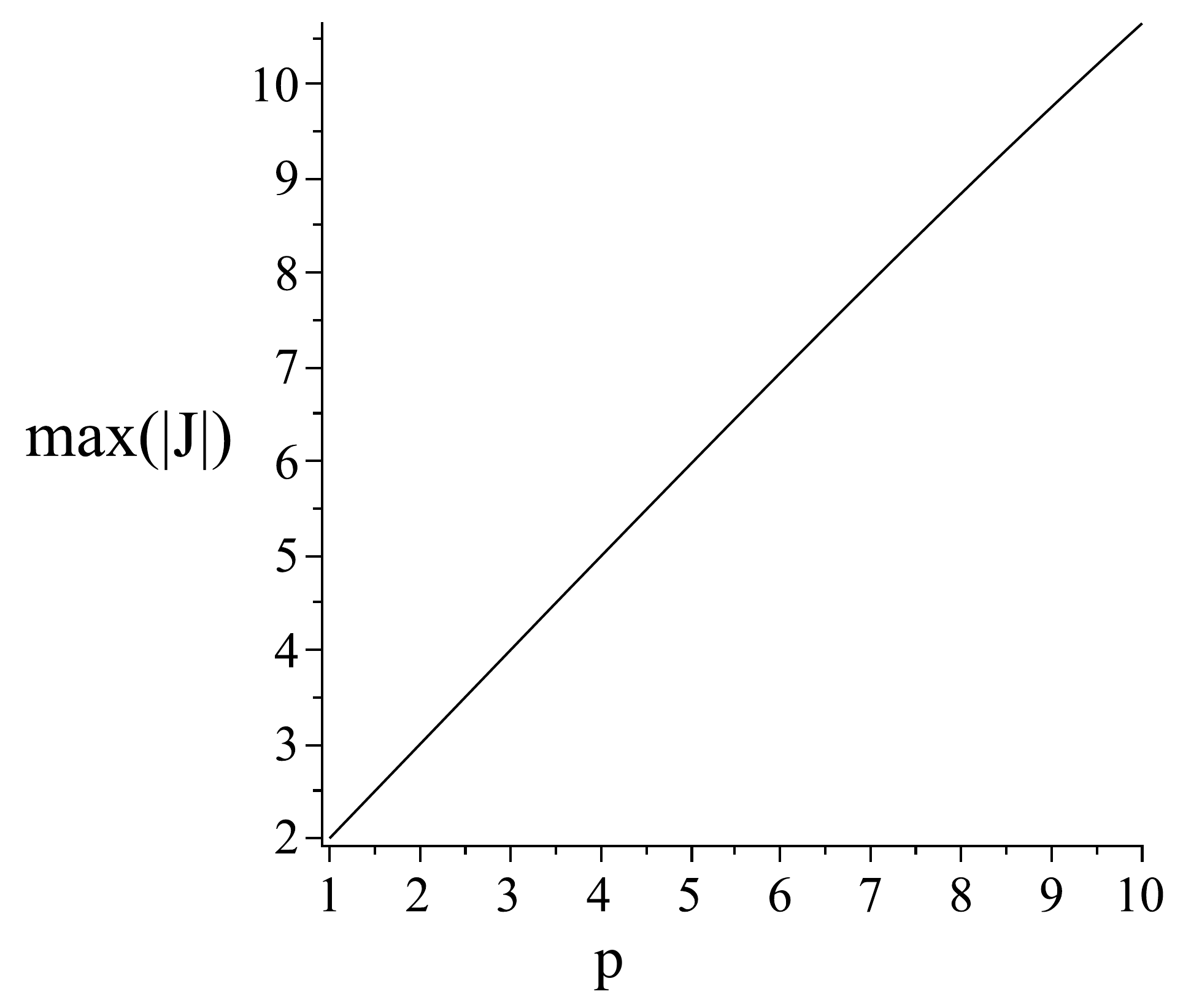}
\includegraphics[width=0.33\textwidth]{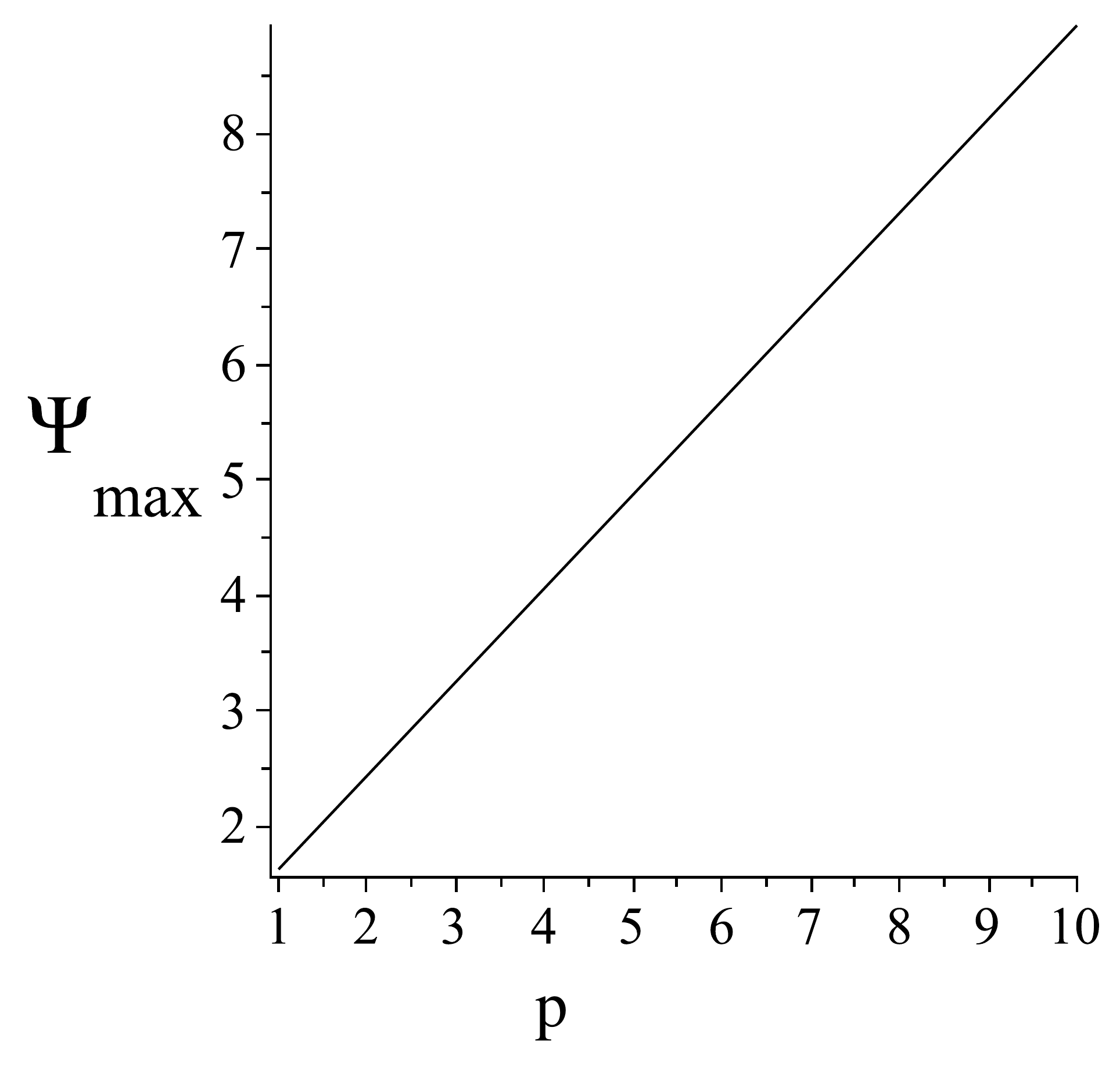}
\caption[]{\label{res_tspine_pdep.fig}  Dependence on the anisotropy parameter $q=p$ of the maximum values of $|{\bf J}|$ and the reconnection rate $\Psi_{max}$ for the kinematic torsional spine model.}
\end{figure}

The results of the above analysis are presented in Figs.~\ref{res_tspine.fig}-\ref{res_tspine_pdep.fig}.
As $p$ is increased,  the current tube shrinks in the $y$-direction, with the dominant current component $J_z$ intensifying in the part of the tube close to the $y$-axis (i.e.~the direction of the short axis of the ellipse) -- see Fig.~\ref{res_tspine.fig}. The stronger current in this region results in an enhanced plasma flow speed. The direction of the flow is also distorted from the circular pattern at $p=1$, but continues to flow on closed elliptical paths around the spine ($z$-axis). As the fan plane is approached the radius of the elliptical shells of positive and negative azimuthal flow increase, owing to the hyperbolic nature of the field structure -- see Fig.~\ref{res_tspine_vdepz.fig}.

In order to determine the reconnection rate we calculate $\Psi$ as defined in Eq.~(\ref{recrate.eq}). Due to the breaking of the symmetry it is no longer clear that the maximal value of $\Psi$ should occur along the spine field line, as was found in previous studies (note that the current modulus has maximum value away from the spine for large $p$). However, it turns out that indeed the maximum occurs along field lines asymptotically close to the spine for all $p$.
Figure \ref{res_tspine_pdep.fig} displays the peak value of the current density (which we impose) and the reconnection rate as a function of the degree of asymmetry. It is clear that the peak current scales linearly with $p(=q)$ and that correspondingly the reconnection rate scales linearly with $p$. Note however that all of the above solutions were obtained with a fixed value of the parameter $j$ which also contributes to controlling the peak current density, and that the velocity and reconnection rate will increase proportional to this parameter. As we shall see below, however, we do not have a linear increase of the peak current with $p$ in our simulations, and it may therefore be more realistic to  take $j=j(p)$.

\subsection{Torsional spine reconnection: simulations}\label{tspine.num}
To complement the above kinematic model we now describe the results of numerical simulations of the full system of resistive MHD equations. The code has been extensively used for other similar simulations, the details of the scheme being described by \cite{nordlund1997}. The MHD equations are non-dimensionalised by setting the magnetic permeability $\mu_0=1$ and the gas constant equal to the mean molecular weight. The result is that for a volume in which $|\rho|=|\BB|=1$, time units are such that an Alfv{\' e}n wave would travel one space unit in one unit of time.
The equations are solved on a numerical grid of $256^3$ gridpoints over $[x,y,z]\in [\pm0.75,\pm0.75,\pm3]$ with uniform $\eta= 10^{-4}$. {The grid is stretched to give higher resolution around the null, where $\Delta x=\Delta y= 0.0034$, $\Delta z=0.014$}. All boundaries are closed and line-tied. We repeat the simulations described by \cite{pontingalsgaard2007}, in which a localised rotational perturbation of the magnetic field is imposed on a background equilibrium null point. Specifically, we begin with a potential magnetic field given by Eq.~(\ref{anis_null.eq}), {plasma velocity zero everywhere, and initialise the plasma density and thermal energy to be spatially uniform with values $\rho=1$ and $e=0.025$, respectively. As a result there exist regions of both high and low plasma-$\beta$ in the domain. $\beta$ approaches infinity as one approaches the null, and takes a value of 1 on an ellipsoidal surface which cuts the coordinate axes at $x=\pm0.09(p+1)$, $y=\pm0.09(p+1)/p$, $z=\pm0.09$.}

In addition to the above, we impose at $t=0$ a magnetic field perturbation composed of a ring of magnetic flux centred on the null point and lying in the fan plane:
\EQ
B_\theta=b_0 \exp\left( 
  - \frac{\left(r-r_{0}\right)^2}{{\alpha}^2} 
- \frac{z^2}{\zeta^2} \right)
\EN
with $b_0=0.05, r_0=0.18, \alpha=0.08, \zeta=0.06$ (see Figs.~1 and 2 of \cite{pontingalsgaard2007}). (We have also performed simulations where this initial perturbation is elliptical rather than circular, but found no change to the qualitative results -- thus here we confine our discussion to the circular perturbation.) 

\begin{table}
\caption[]{\label{tspine.tab} Data on the simulations of torsional spine reconnection.} 
\begin{tabular}{ c c c c c c r}
\hline
\hline
$p$  & ${|{\bf J}|_{max}}$  & $(J_z>0)_{max}$& ${\Psi_{max}}^{~\mathrm{a}}$ & $L_x^{~~\mathrm{b}}$ & $L_y^{~~\mathrm{b}}$ & $L_z^{~~\mathrm{b}}$\\ 
\hline 
1 & $0.31$  &   0.24& $4.2\times 10^{-5} $    &0.03 & 0.03      & 3.9     \\ %
2 & $0.29$  &   0.27& $5.7\times 10^{-5} $    & 0.09& 0.023        &  3.9    \\ %
3 & $0.31$  &  0.31&  $6.0\times 10^{-5} $  & 0.13 & 0.019        &  3.3    \\ %
5 & $0.34$  & 0.34&  $6.1\times 10^{-5} $   & 0.19 & 0.018        &  2.5    \\ %
10 & $0.36$  & 0.36&  $5.8\times 10^{-5} $    & 0.26 & 0.017       &  1.9    \\ %
\hline
\end{tabular}
\begin{list}{}{}
\item[$^{\mathrm{a}}$] peak integrated parallel electric field attained.
\item[$^{\mathrm{b}}$] current layer dimensions measured at the time when $J_z>0$ reaches its temporal maximum.
\end{list}
\end{table}

\begin{figure} \centering
 \includegraphics[width=0.49\textwidth]{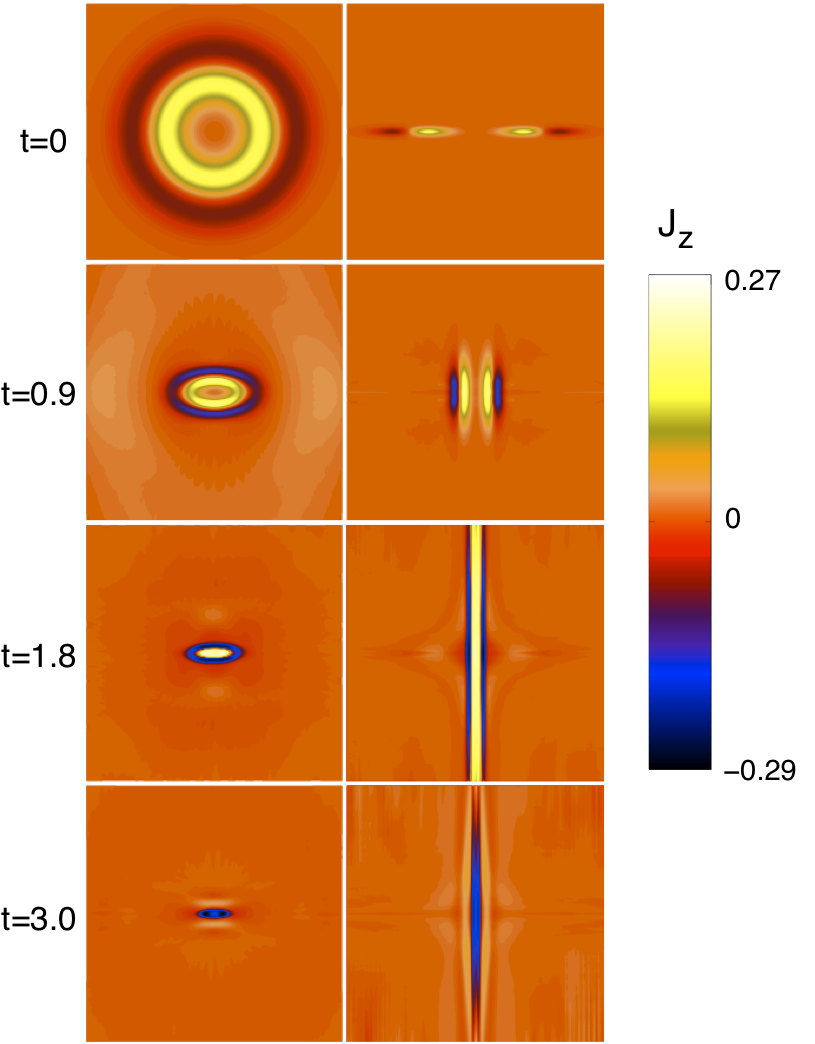}
\caption[]{\label{j_tspine_seq.fig} Time sequence showing contours of $J_z$ in the $z=0$ (left) and $x=0$ (right) planes, for the torsional spine simulation with $p=2$.}
\end{figure}

\begin{figure} \centering
\centering
  \includegraphics[width=0.16\textwidth]{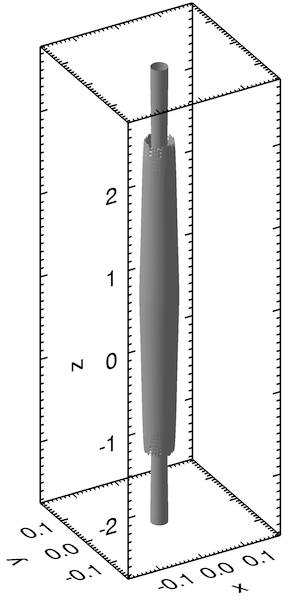}
 \includegraphics[width=0.16\textwidth]{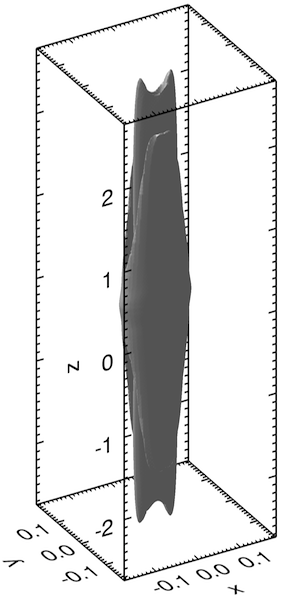}
 \includegraphics[width=0.16\textwidth]{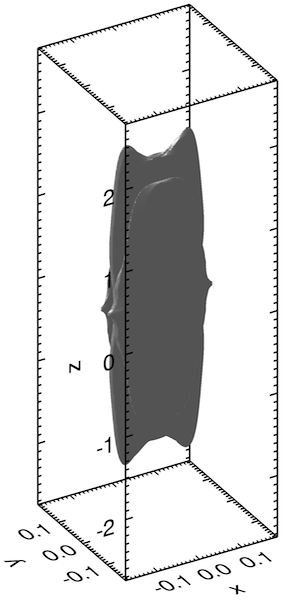}\\
 \includegraphics[width=0.5\textwidth]{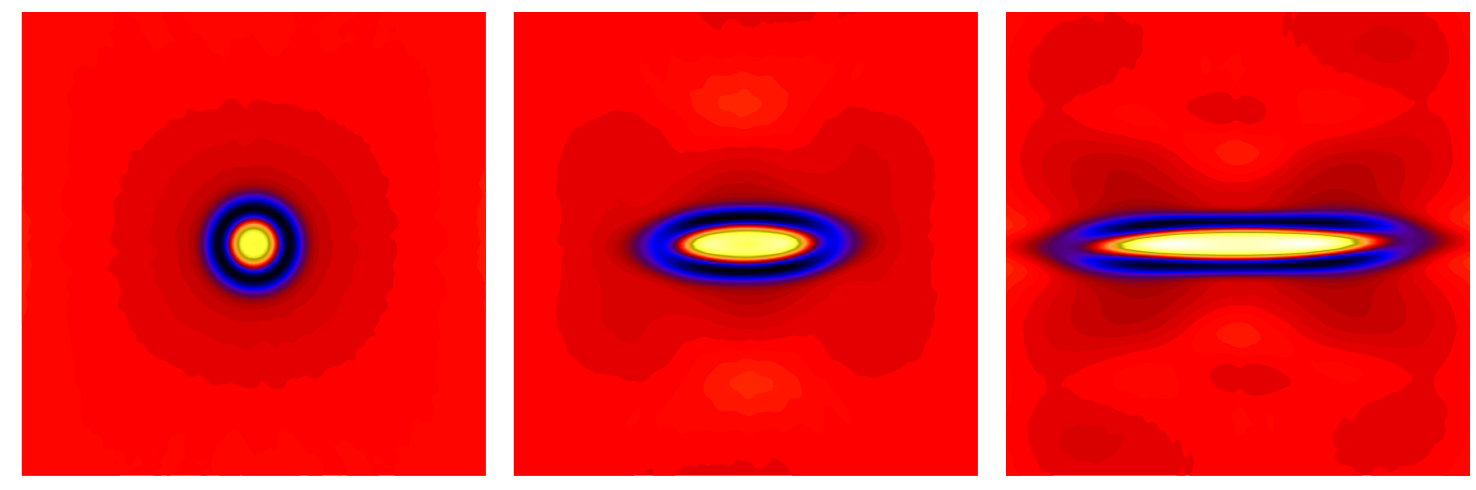}\\
 \includegraphics[width=0.25\textwidth]{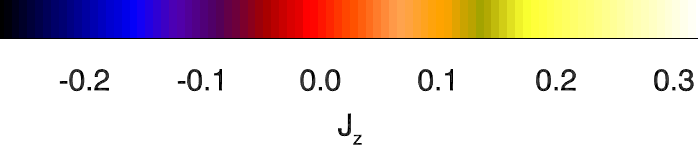}
\caption[]{\label{jiso_tspine.fig} Above: Current isosurface at 50\% of maximum value, for the torsional spine simulations with $p=1$ (left), $p=2$ (middle) and $p=5$ (right). Taken in each case at the time when the positive value of $J_z$ reaches a maximum. Below: contours of $J_z$ in the $z=0$ (fan) plane over $[x,y]\in[\pm0.2,\pm0.2]$ for the same runs at the same times.}
\end{figure}

For $t>0$ the perturbation splits, with wavefronts travelling both toward and away from the null. We focus on the behaviour of the ingoing pulse, which for $p=1$ gradually stretches out to form a cylindrical tube of intense current around the spine. When $p\neq 1$, the azimuthal symmetry of the perturbation wavefront is broken as soon as the evolution begins. The Alfv{\' e}n speed in the radial direction now depends on $x$ and $y$, and so the wavefront travels toward the spine faster along the $y$-axis where $|\BB|$ is stronger. Propagation in the $z$-direction is essentially unaffected, and the current distribution associated with the perturbation forms into a cylinder with elliptical cross-section, whose length and eccentricity both increase as the pulse steepens towards the spine and the peak current correspondingly intensifies (see Figs.~\ref{j_tspine_seq.fig}, \ref{jiso_tspine.fig}).

\begin{figure} \centering
 \includegraphics[width=0.4\textwidth]{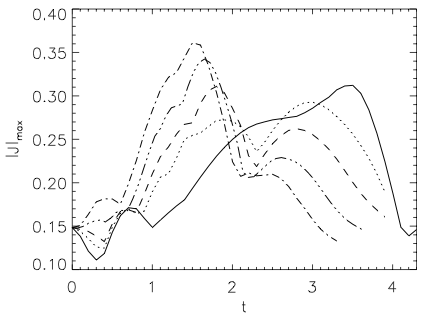}
\caption[]{\label{j_tspine.fig} Evolution of the peak value of $|\JJ|$ for the different torsional spine simulations: $p=1$ (solid line), $p=2$ (dotted), $p=3$ (dashed), $p=5$ (triple-dot-dashed) and $p=10$ (dot-dashed).}
\end{figure}

As shown in Fig.~\ref{j_tspine.fig}, there is a strong (at least 100\%) increase in the peak current density as the disturbance reaches the spine. It is also clear from the plot that this occurs earlier for larger $p$ due to the increased speed of propagation along the $y$-axis. In general, the peak current is higher for larger $p$. However, this is complicated by a competing effect -- namely that there are two distinct maxima in $|\JJ|_{max}$ during the simulations, one corresponding to the localisation of the leading edge of the pulse and the other to the trailing edge of the pulse. As can be readily seen in Fig.~\ref{j_tspine_seq.fig}, they correspond to opposite signs of $J_z$. Thus, while the maximum positive value of $J_z$ strictly increases as $p$ increases, it is found that for $p=1,2$ the trailing edge of the pulse (corresponding to $J_z<0$) dominates. The dimensions of the current layer are shown in the final three columns of Table \ref{tspine.tab}. They are measured at a time corresponding to the localisation of the leading edge of the pulse, i.e.~the time when $J_z>0$ reaches its temporal maximum (which occurs at a time marked by the peak on each curve in Fig.~\ref{j_tspine.fig} between $t=1$ and $t=2$, except for the run with $p=1$ in which $J_z$ reaches its maximum at $t=2.1$). The dimensions of the current tube in the $xy$-plane demonstrate its elliptical nature (centred on the spine) with eccentricity increasing with $p$ as shown in Fig.~\ref{jiso_tspine.fig}. 
The length in the $z$-direction ($L_z$) decreases with increasing $p$. This is because the $z$-component of $\BB_P$ (and thus the propagation speed in $z$) is independent of $p$, and because the pulse localises in the $xy$-plane at an earlier time for larger $p$ as explained above.
It is also worth noting from Fig.~\ref{jiso_tspine.fig} that for $p\neq 1$ the maximum current is attained not exactly on the spine, but in two locations displaced symmetrically from the spine along the $x$-axis.

\begin{figure} \centering
 \includegraphics[width=0.4\textwidth]{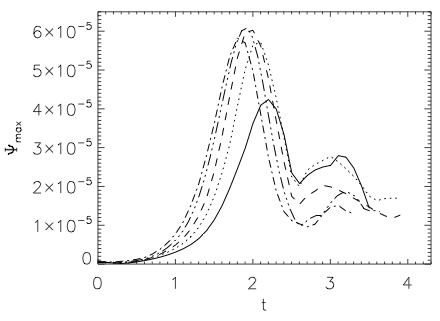}
\caption[]{\label{recrate_tspine.fig} Evolution of the reconnection rate for the different torsional spine simulations: $p=1$ (solid line), $p=2$ (dotted), $p=3$ (dashed), $p=5$ (triple-dot-dashed) and $p=10$ (dot-dashed).}
\end{figure}
We turn now to consider the reconnection rate, calculated as described in Eq.~(\ref{recrate.eq}).  The maximal value of $\Psi$ is found over all field lines that thread the current layer, these being field lines that pass close to the null and its spine and fan. For $p=1$, due to symmetry, the maximum can be found on any field line which runs asymptotically close to the null. However, for $p>1$ the current is strongest in the weak field regions around the $x$-axis, and so field lines that pass through these regions attain the highest values of $\Psi$. The evolution of the reconnection rate is plotted for the different simulations in Fig.~\ref{recrate_tspine.fig}. The maximum value  attained does not depend strongly on $p$ (see the fourth column of Table \ref{tspine.tab}), except that it is significantly lower for $p=1$. However, this is likely down to the fact that for $p=1$ the disturbance extends all the way to the $z$-boundaries, and so we `miss' some of the length of the current layer. Note that the reconnection rate calculated here is a net effect of integrating through regions of both positive and negative $\EE\cdot\BB$ --  we have different senses of reconnection (rotational slippage) occurring on the leading and trailing edges of the pulse, and here we measure the net effect. 

\subsection{Torsional spine reconnection: discussion}\label{tspine.discuss}
The numerical simulations discussed above demonstrate the localisation of a rotational perturbation towards the spine of a non-symmetric linear 3D null. The resulting current intensification is associated with torsional spine magnetic reconnection. The current tube that forms around the spine has a structure that is closely matched by the kinematic steady-state model described in Sect.~\ref{tspine.kin}. In particular, the current is dominated by the component parallel to the spine ($J_z$), and is localised within a tube of elliptical cross-section. The short axis of the ellipse is aligned with the strong field direction in the fan plane, along which the current is most intense. The eccentricity of the ellipse increases as the magnetic field asymmetry increases. {This dynamic evolution should be compared with the stationary analytical solution at the stage where the current layer has become fully localised on the spine, as in the analytical solution -- i.e.~when the maximum (or minimum) of $J_z$ is achieved. If it were feasible to drive the perturbation from the boundaries in a steady fashion, we would expect this spine-localised current tube to persist in a quasi-steady state for a period controlled by the period of steady driving, as in Sect.~\ref{tfan.sec} below -- see Fig.~\ref{jmax_sprtd.fig} (driving a rotational motion in a smooth way around the fan boundaries in our Cartesian domain is not practical).}

The plasma flow also has a similar qualitative structure in the kinematic model and simulations. This structure is that of a rotation along elliptical paths around the spine, with these elliptical paths closely following the current density contours. 
One difference between the model and simulations is that {both signs of rotational flow are only seen during a certain period in the simulations. Specifically, as the twist associated with the perturbation propagates towards the null it drives flow predominantly in the positive rotational sense, and when the reconnection process is completed ($t \gtrsim 3.5$) the field then `untwists' leading to a large-scale flow in the opposite direction. It is only approximately between the times that $J_z$ reaches its maximum positive and negative values (approximately $1.5<t<3$, see Fig.~\ref{j_tspine.fig}) that the Lorentz force accelerates the plasma in opposite rotational senses at different distances from the null.}
By contrast, in the steady model where momentum balance is neglected, rotational plasma flows of both senses are required to maintain the steady state.

In the numerical simulations it is found that the peak reconnection rate is approximately independent of $p$.
We can understand why this should be the case by noting that, while $J_z$ increases with increasing $p$, the length of the current layer along $z$ decreases with increasing $p$. Thus for larger $p$ we have a larger integrand in Eq.~(\ref{recrate.eq}), but it is non-zero over a shorter distance. Note also that the reconnection rate is found to be given by the integral of $E_\|$ along field lines lying asymptotically close to the spine. Since ${\bf J}$ is dominated by $J_z$ and the current layer has minimal extent in the $xy$-plane, then it is natural that the reconnection rate should not depend strongly on the ${\bf B}_{xy}$ field components away from the null. In order to match these simulation results for the $p$-dependence of the reconnection rate, the magnetic field ${\bf B}_J$ in our kinematic model could be normalised by a factor proportional the current modulus.

It is worth considering here the interpretation of the reconnection rate as measured above. Within the framework of general magnetic reconnection \citep{schindler1988}, the flux reconnected is measured by calculating the differential transport of flux on opposite sides of the non-ideal region, for a flux surface passing through the peak of the pseudo-potential $\Psi$ \citep[see also][]{hesse2005}. This theory is based on the assumption that no nulls are present within the non-ideal region, while here of course we consider a non-ideal region containing a null point.
In the rotationally symmetric case the maximal value of $|\Psi|$ is found on any field line passing asymptotically close to the null. This measures the differential rate of flux transport through any flux surface bounded by the spine and fan in the half-space (above or below the fan), by symmetry  \citep{pontin2004,pontin2011b}. That is, it measures the net  magnetic flux reconnected at the given time. However, in the present simulations for $p\neq 1$ the value of $\Psi$ no longer exhibits azimuthal symmetry. Therefore the maximal value of $|\Psi|$ obtained above measures the maximum differential rate of flux transport for any flux surface bounded by the spine and fan, though one should bear in mind that this now varies around the spine.
Of course in a steady-state situation, this differential rate of flux transport must be the same for all such flux surfaces (by continuity), and indeed it is found to be in the steady-state kinematic solution.

The behaviour of the disturbance -- spreading out along the spine in all of the simulations -- suggests that the dominant wave mode is an Alfv{\' e}n wave. However, it is highly likely that other wave modes are present. Owing to the differing waves speeds approaching the null for different directions, one may speculate that if the disturbance were sufficiently strong as to be considered non-linear, then some of the effects discussed by \cite{mclaughlin2009} would be present. Those authors considered the propagation of a non-linear fast-mode wave towards a 2D X-point, and observed the formation of cusp-shaped structures as the wavefront collapsed, and the formation of both fast and slow mode shock waves. However, since our simulation is three-dimensional, it is not feasible at present  to use the numerical resolution required to properly resolve such features. {Note further that the perturbation in our simulations is initialised in the vicinity of the $\beta=1$ surface. It is known that wave-mode conversion may take place in the vicinity of this surface. Thus the propagation of the disturbance towards the null, as well as the final form of the current layer, may be different if we were to repeat the series of simulations with different values of the plasma-$\beta$.}

\section{Torsional fan reconnection}\label{tfan.sec}
\subsection{Torsional fan reconnection: kinematic model}\label{tfan.kin}

\begin{figure} \centering
 \includegraphics[width=0.4\textwidth]{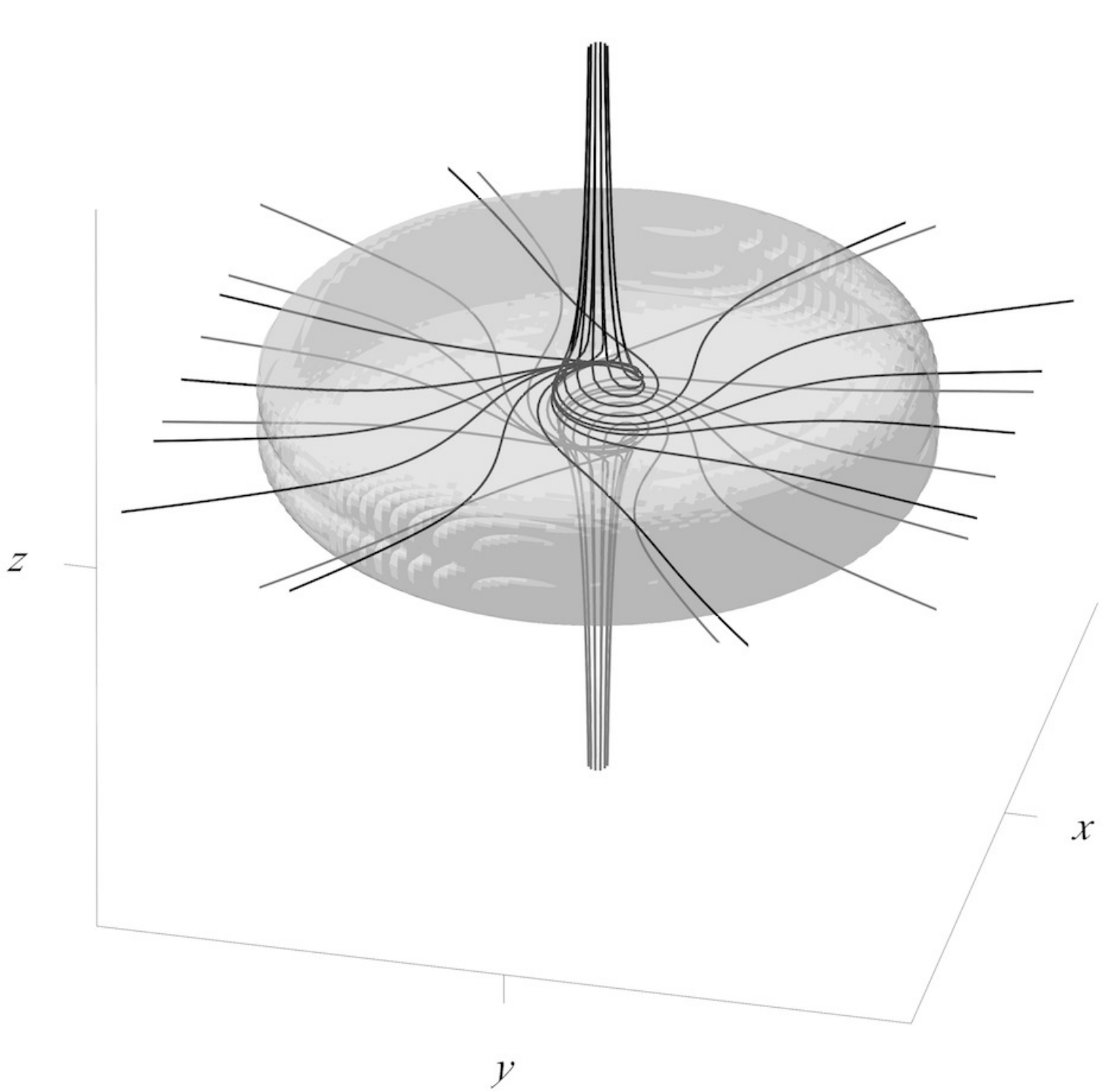}
 \includegraphics[width=0.4\textwidth]{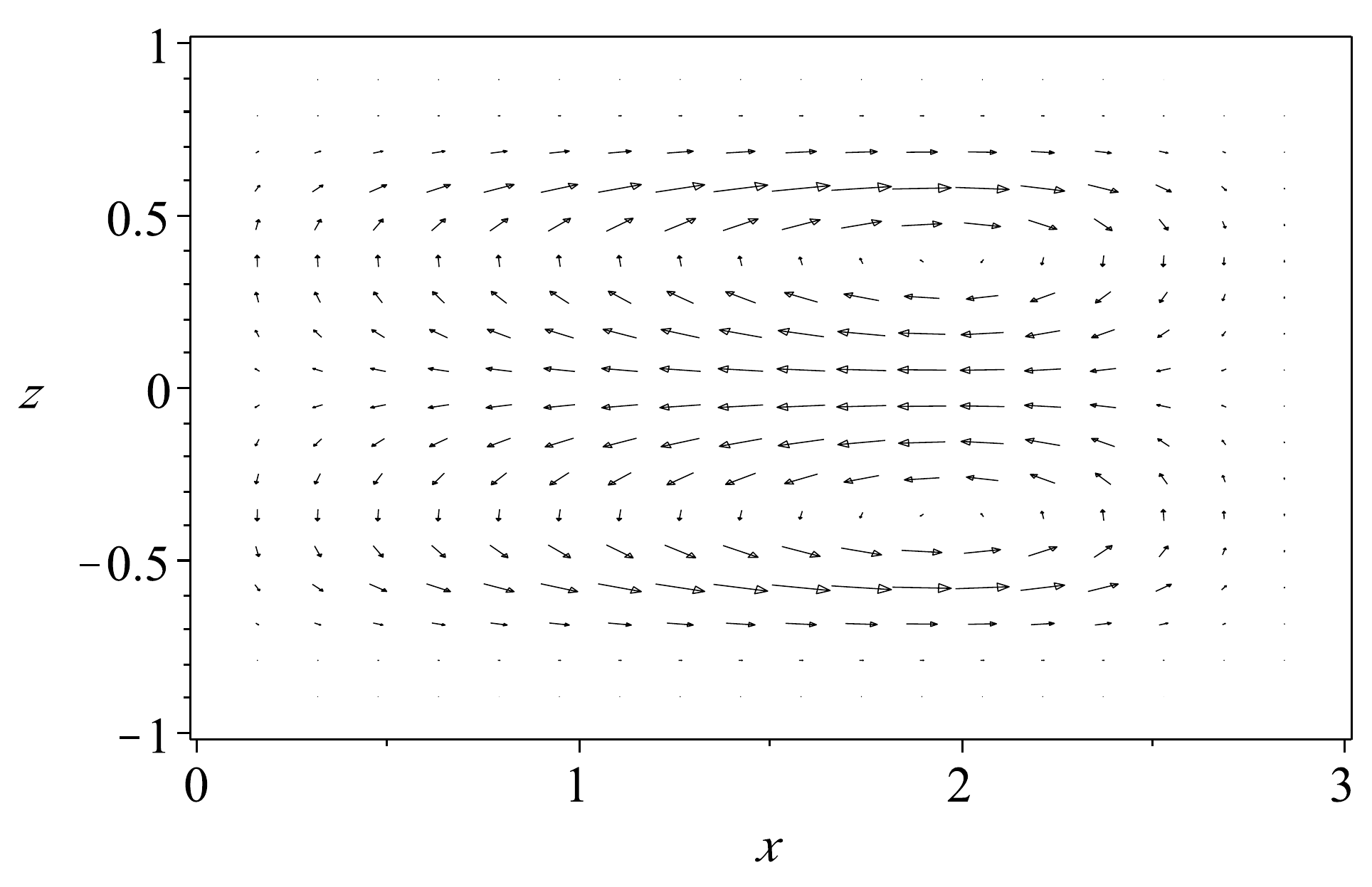} 
\caption[]{\label{kin_tfan.fig} {\it Above}: Magnetic field lines for the torsional fan model defined by Eqs.~(\ref{totb.eq},\ref{anis_null.eq},\ref{anis_fanb.eq}) for $a=5,b=1,j=50,p=1.5,q=1$. The shaded surface shows a current isosurface. {\it Below}: current vectors in the $y=0$ plane for $a=3,b=1,q=1$.}
\end{figure}

We now turn our attention to modelling of the torsional fan reconnection mode, which involves rotational slippage of field lines in a current layer localised around the fan surface. We proceed to solve Eqs.~(\ref{kineq.eq}) in the same way as described in Sect.~\ref{tspine.kin}. Again, we first analyse a model for the cylindrically symmetric case, in which for the first time a localised current layer is included in the fan plane. The structure of the magnetic field is chosen by comparing with the  numerical simulations of \cite{galsgaardpriest2003,pontingalsgaard2007}. We again construct our magnetic field as the sum of a potential part ($\BB_P$) and non-potential part ($\BB_J$) as in Eq.~(\ref{totb.eq}), with $\BB_P \!=\! [r, 0, -2z]$, and this time
\EQ
\BB_J \!=\! \left\{ 
\begin{array}{cc}
\left[ 0, jrz \left(1-\left(\frac{r}{a}\right)^{2m}\right)^{2\mu}\left(1-\left(\frac{z}{b}\right)^{2n}\right)^{2\nu},0 \right] &~~~ 
\begin{array}{c}
r\leq a  \\ \&~ |z|\leq b
\end{array}
 \\
 \rule{0pt}{3ex} 
\left[0,0,0 \right] & ~~~ {\rm otherwise,}
\end{array}
\right. \label{fanb.eq}
\EN
see Fig.~\ref{kin_tfan.fig}. Note that $B_\theta$ is now odd in $z$, and since we are modelling a current layer focused on the fan plane we assume that $b \ll a$. We again choose the integers $m,n,\mu,\nu$ in such a way that all physical quantities in our solution are continuous and differentiable, specifically $m=3, \mu=n=2, \nu=6$. The field line mapping for $\BB$ is again given by  $r=r_0 \exp(B_0 s/L_0)$ and $z=z_0 \exp(-2B_0s/L_0)$, along with a lengthy expression for $\theta(r_0,\theta_0,z_0,s)$ which is not required to obtain the solution. This time we solve Eqs.~(\ref{kinsol}) setting $s=0$ at $r=r_0=a$. We choose to do this because it leads to non-zero flow for $|z|>|b|$, which is consistent with the observed result from the simulations that this reconnection mode is set up by rotational driving flows in the regions around the spine footpoints.

\begin{figure} \centering
\includegraphics[width=0.34\textwidth]{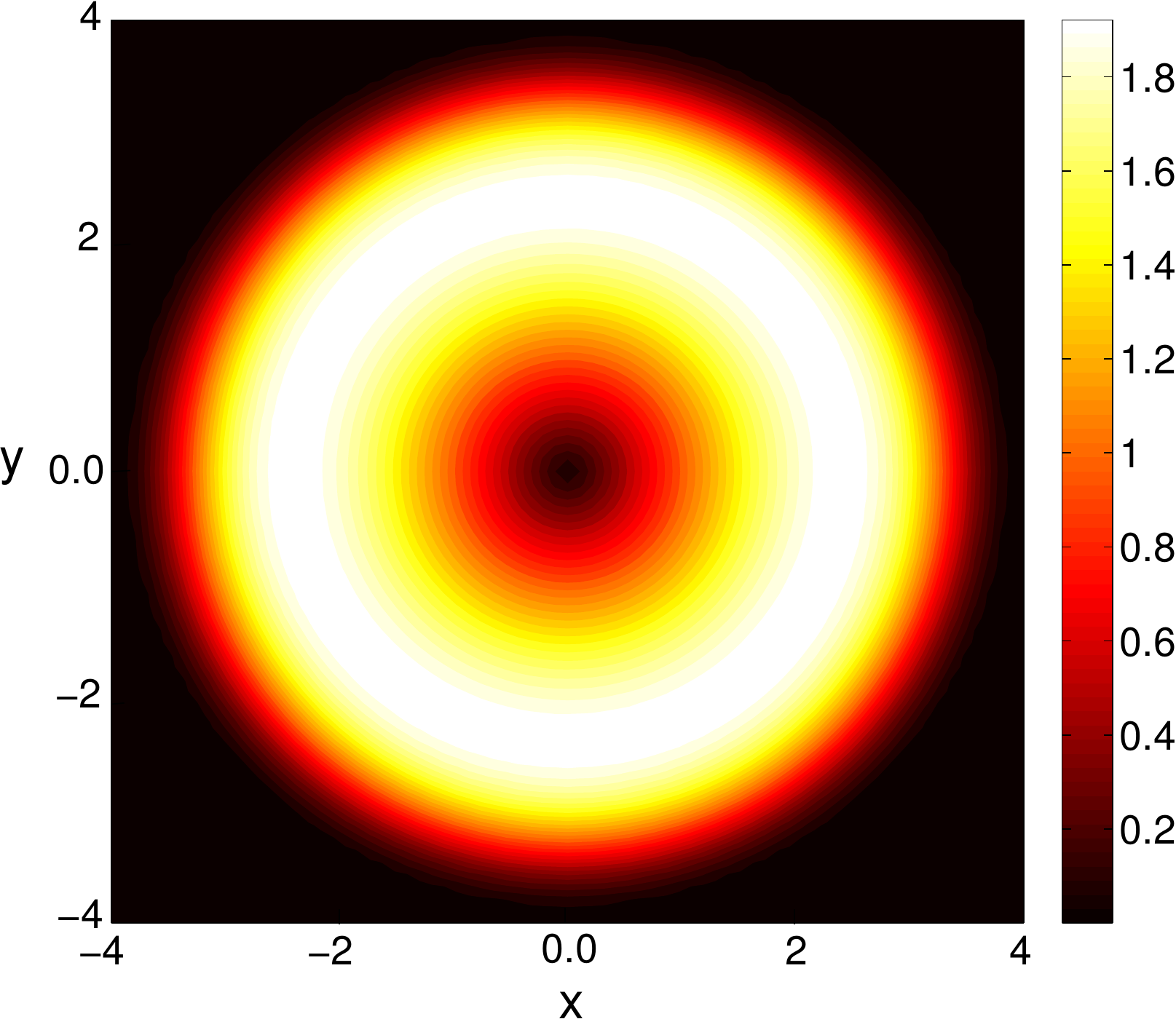}
\includegraphics[width=0.34\textwidth]{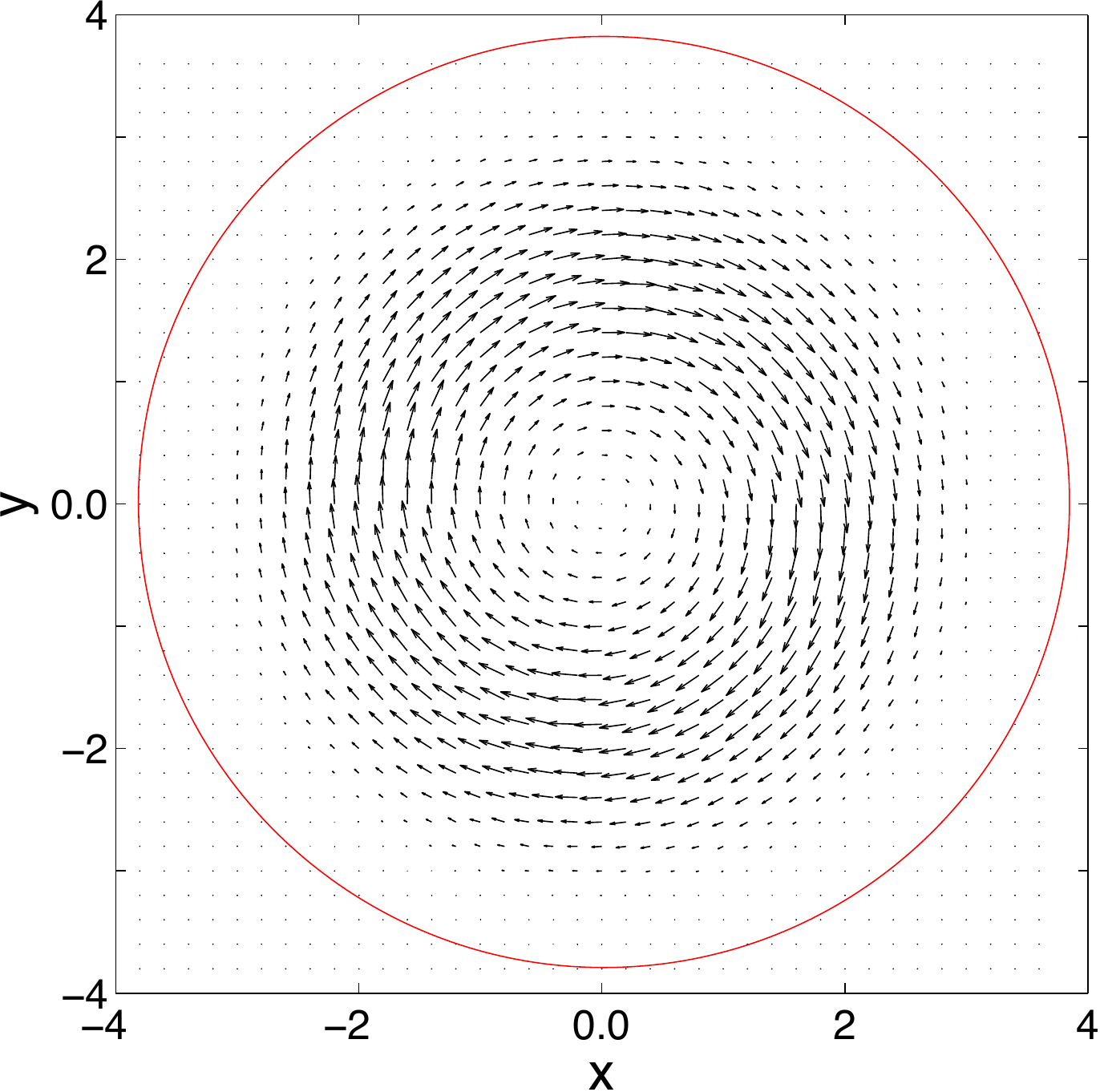}\\
\includegraphics[width=0.34\textwidth]{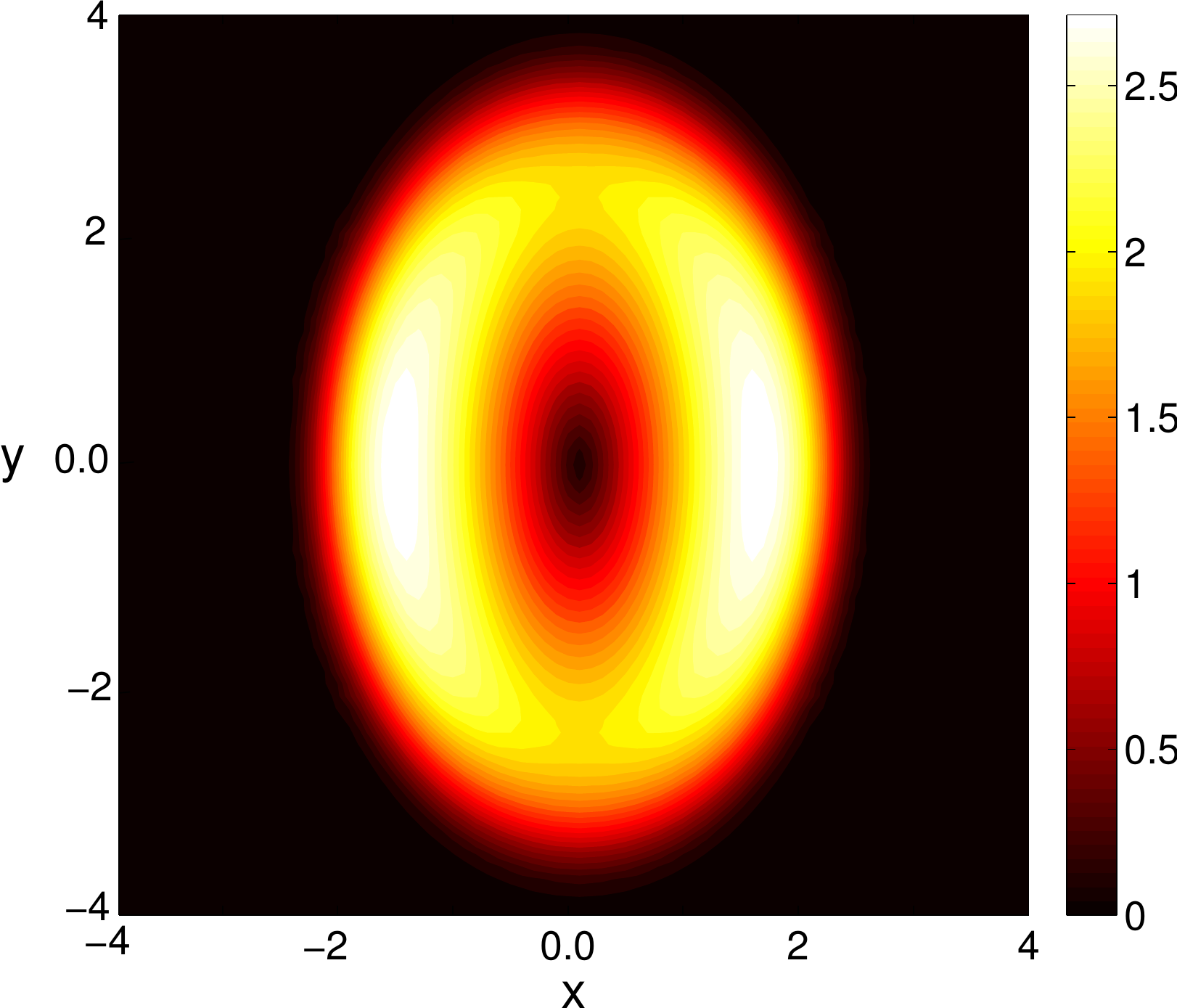}
\includegraphics[width=0.34\textwidth]{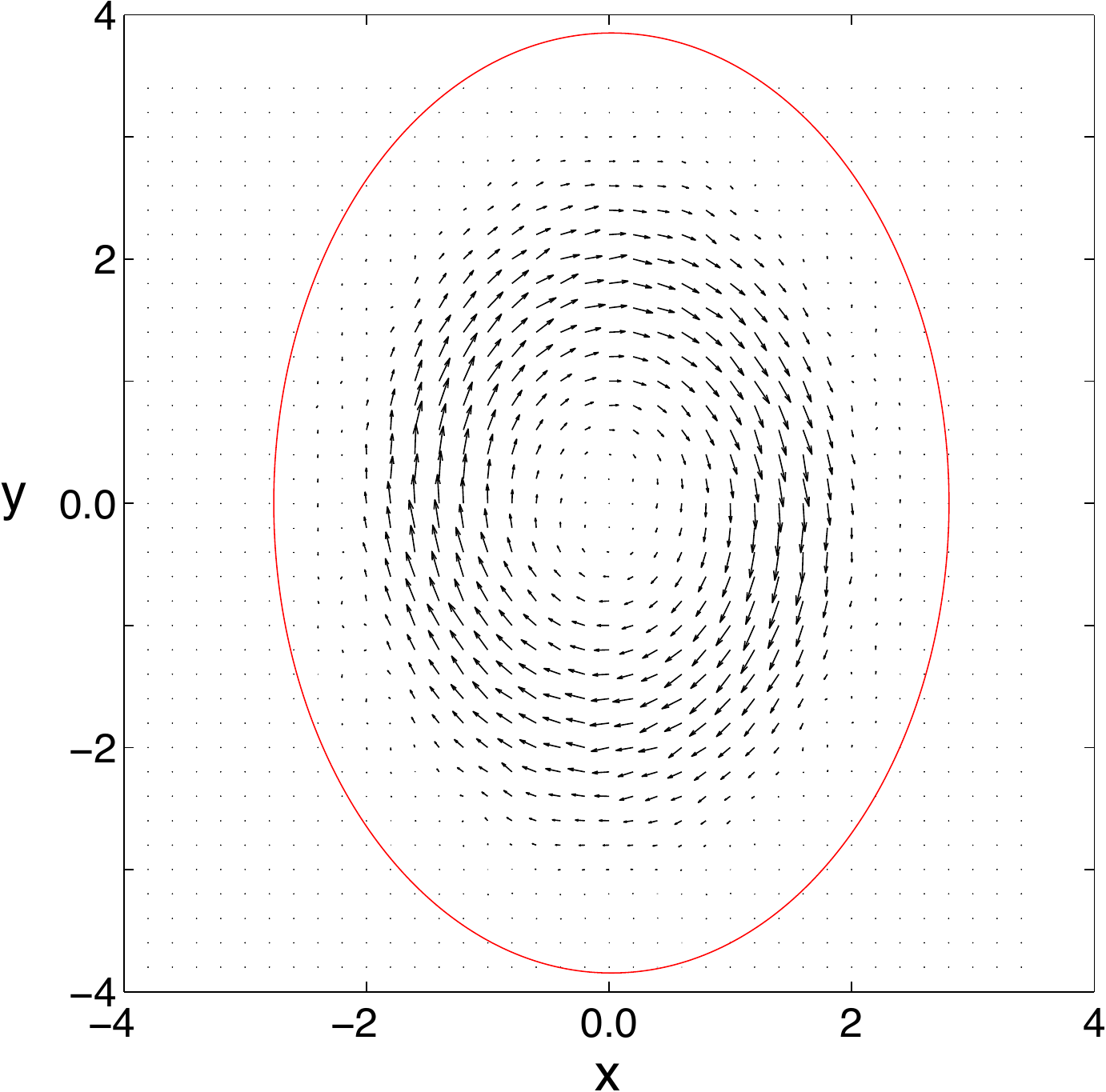}\\
\includegraphics[width=0.34\textwidth]{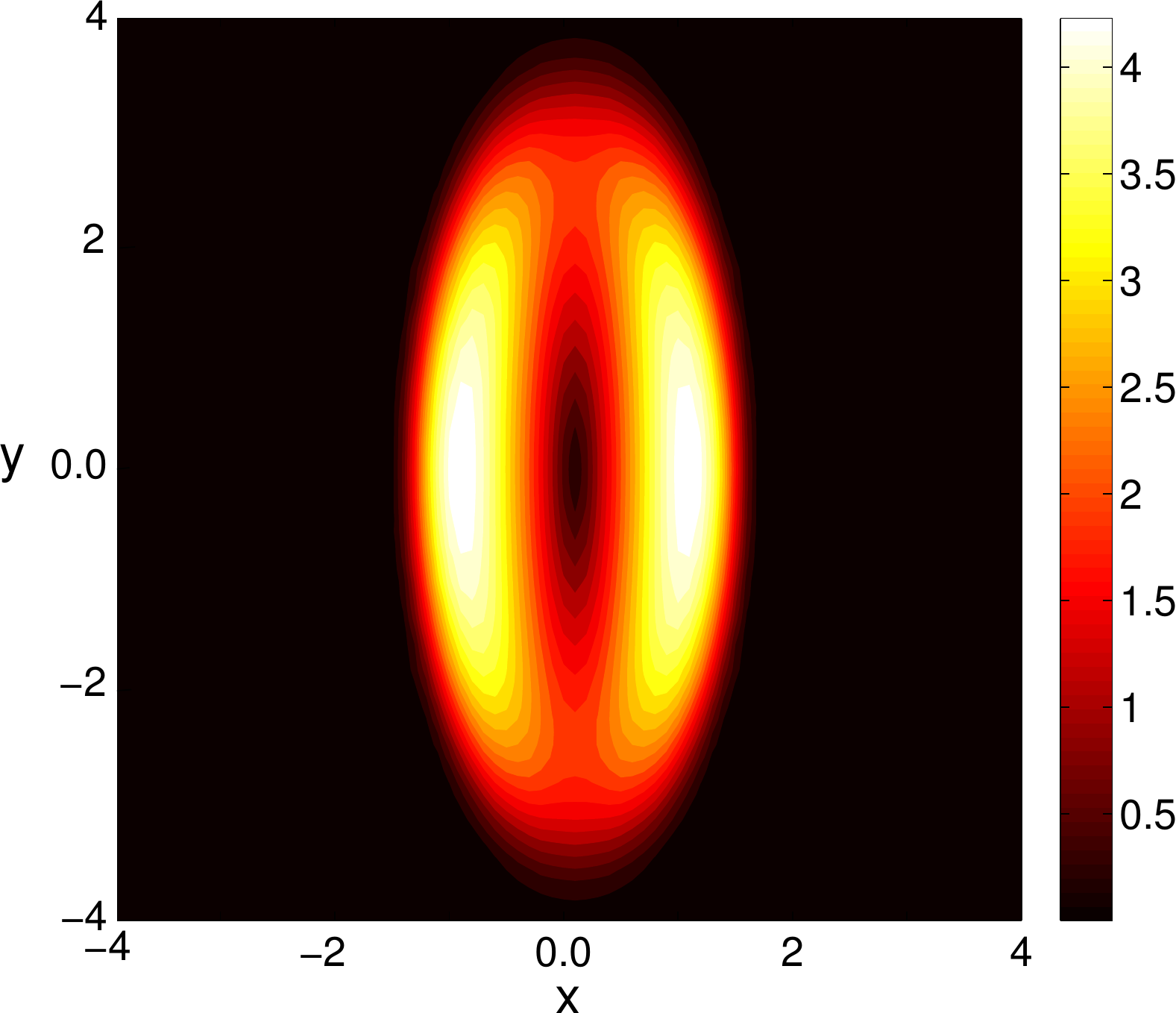}
\includegraphics[width=0.34\textwidth]{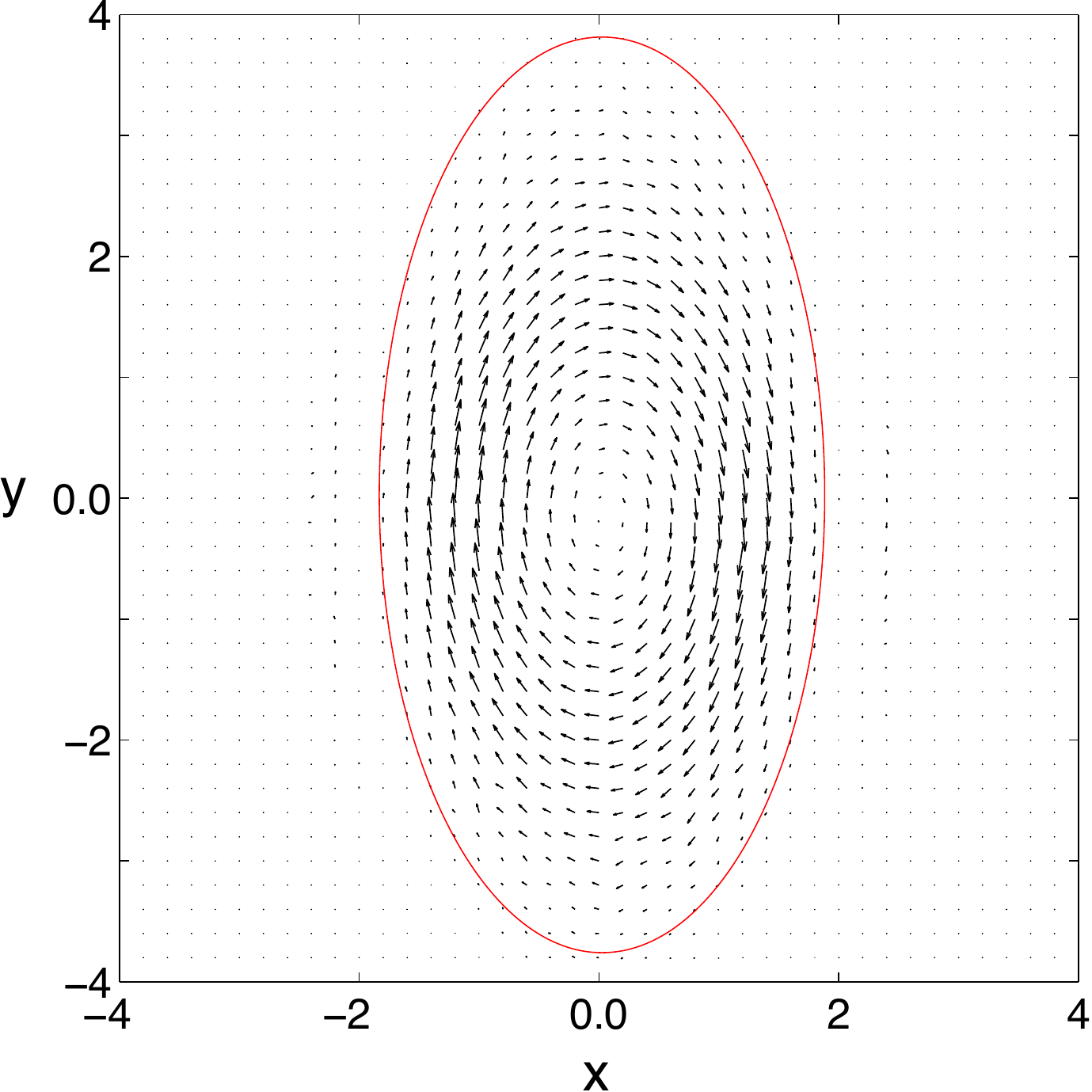}
\caption[]{\label{res_tfan.fig} Current density (left) and plasma flow vectors (right) in the kinematic torsional fan model. Plotted for $B_0=L_0=j=b=1$, $a=4$ and $p=q=1$ (above), $p=q=2$ (middle) $p=q=5$ (below). The grey curve (red online) in the right-hand plots marks the boundary of the non-ideal region. The current density is plotted at $z=0$, while the plasma flow is plotted in the plane $z=0.5$.}
\end{figure}
\begin{figure} \centering
\includegraphics[width=0.34\textwidth]{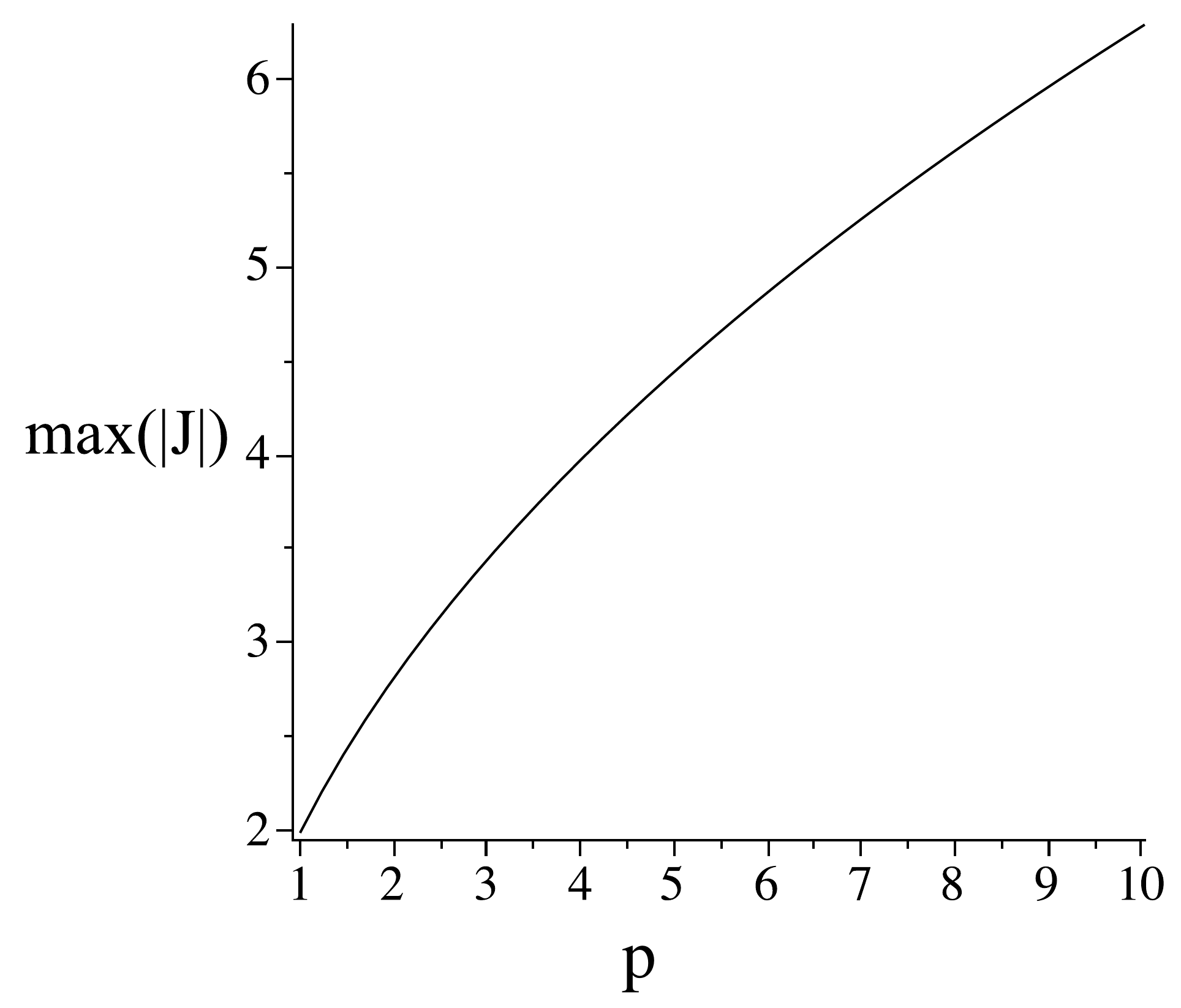}
\includegraphics[width=0.34\textwidth]{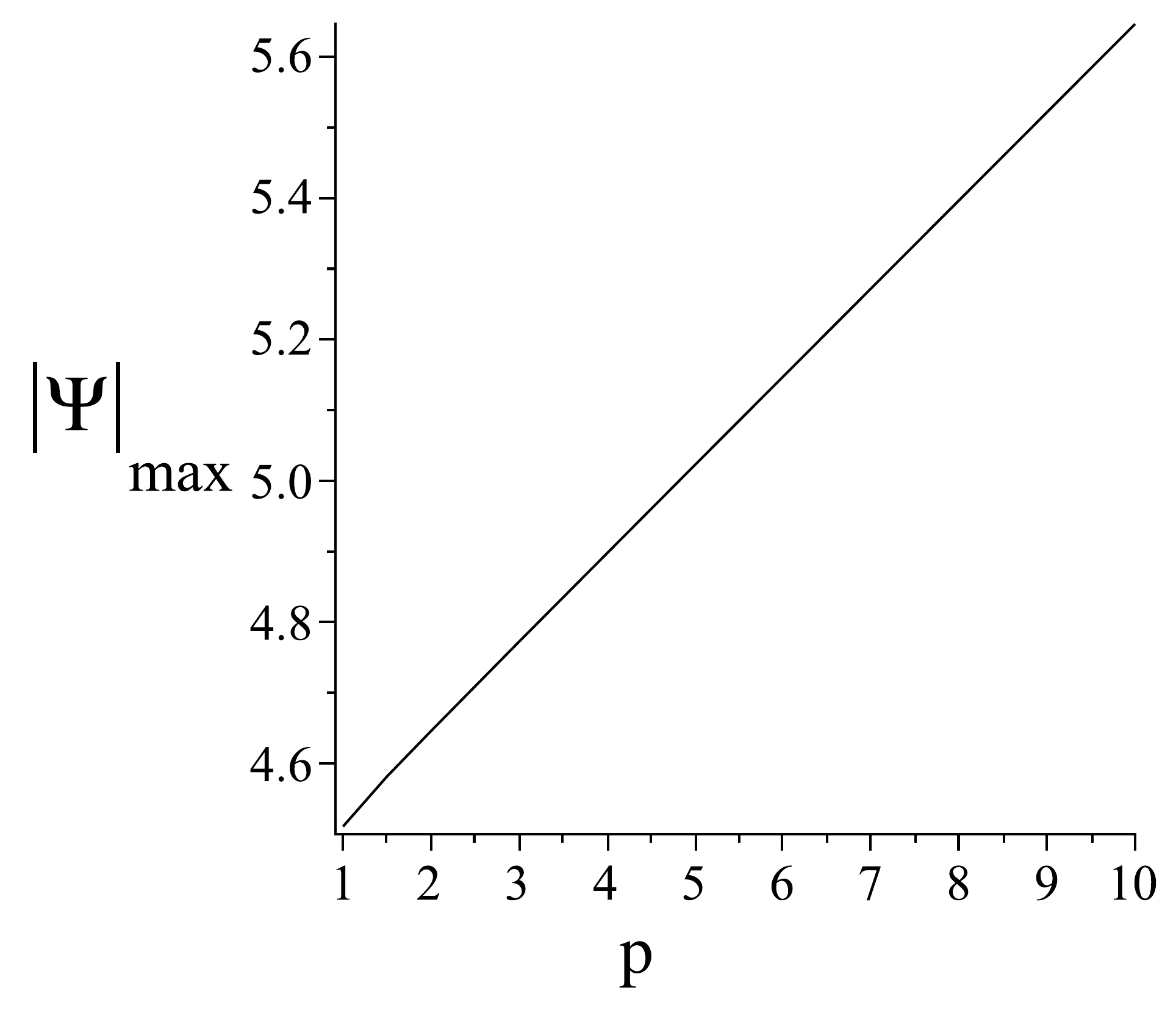}
\caption[]{\label{res_tfan_pdep.fig} Dependence on the anisotropy parameter $q=p$ of the maximum values of $|{\bf J}|$ and the reconnection rate $\Psi_{max}$ for the kinematic torsional fan model. }
\end{figure}

The new torsional fan solution is represented in Figs.~\ref{res_tfan.fig}, \ref{res_tfan_pdep.fig}. The  plasma flow has a rotational structure, as in the solution of \cite{pontin2004}. That is, when we subtract a component of $\vv$ parallel to $\BB$ such that $v_z=0$, then the remaining flow is non-zero only in the azimuthal direction (see the top-right frame of Fig.~\ref{res_tfan.fig}). Thus, field lines traced from comoving footpoints in the ideal region at $z>z_0$ (or $z<-z_0$) rotate around the spine at a fixed radius, while field lines traced from the ideal region at $r>r_0$ remain fixed ($\vv=0$ there), and we have a change of connectivity in the form of a counter-rotational slippage. Owing to the fact that $J_r$ changes sign for different levels of $z$, the rotational flow within the current layer has regions of both clockwise and anti-clockwise rotation, in a similar way to the torsional spine model discussed above. 

As before, we  now  investigate the dependence of the properties of our solution on the symmetry of the magnetic field. We proceed  as in Sect.~\ref{tspine.kin} to solve Eqs.~(\ref{kinsol}) using the semi-analytical method described there. 
$81^3$ gridpoints are used over the domain $-4 \leq x,y \leq 4$, $0\leq z \leq 2$, and we use parameters
$B_0=L_0=j=\eta_0=1$, $b=1, a=4$.

We take the potential component of our magnetic field ($\BB_P$) to be given by Eq.~(\ref{anis_null.eq}), with the non-potential component taken to be
\EQ
\BB_J = 
 \left\{ 
\begin{array}{cc}
jz \left(1-\left(\frac{R}{a}\right)^{6}\right)^{4}\left(1-\left(\frac{z}{b}\right)^4\right)^{12}
\left[ -y, qx ,0 \right] &~~~ 
\begin{array}{c}
R\leq a  \\ \&~ |z|\leq b
\end{array}
 \\
 \rule{0pt}{3ex}
\left[0,0,0 \right] & ~~~ {\rm otherwise}
\end{array}
\right. \label{anis_fanb.eq}
\EN
where $R^2=qx^2+y^2$ (which reduces to Eq.~(\ref{fanb.eq}) when $q=1$). 
The current layer now has the shape of an elliptical disc, with major and minor axes along the $x$- and $y$-axes, extending to $x=\pm a/\!\sqrt{q}$, $y=\pm a$. 
Pre-empting the results of the following section, we present here results for $p=q$, such that as $p$ increases the current disc shrinks along the direction associated with the small fan eigenvalue, i.e.~the weak field direction in the fan. As shown in Fig.~\ref{res_tfan.fig}, the current density is enhanced in the regions around the short axis of the ellipse. As in the torsional spine solution, if we set the parallel flow in such a way as to eliminate $v_z$, then the plasma flow in the $xy$-plane follows closed elliptical paths, being strongest in magnitude where the current is enhanced. 

Examining the dependence of the peak current and reconnection rate on the degree of asymmetry, we find that both increase with increasing $q$, as shown in Fig.~\ref{res_tfan_pdep.fig}.

\subsection{Torsional fan reconnection: simulations}\label{tfan.num}

\begin{table}
\caption[]{\label{tfan.tab} Data on the simulations of torsional fan reconnection.} 
\begin{tabular}{ c c c} 
\hline
\hline
$p$  & ${|J_{xy}|_{max}}$  & ${\Psi_{max}}$ \\ 
\hline 
1 & $0.94$  &    $12.7\times 10^{-5} $   \\ 
2 & $1.20$  &    $7.4\times 10^{-5} $   \\ 
3 & $1.33$  &    $5.7\times 10^{-5} $  \\ 
5 & $1.49$  &   $4.5\times 10^{-5} $  \\ 
10 & $1.55$  &  $3.8\times 10^{-5} $   \\ 
\hline
\end{tabular}
\end{table}

We now perform numerical simulations similar to those described in Sect.~\ref{tspine.num}, designed to investigate the effect of the background field asymmetry on the torsional fan reconnection mode (see Table \ref{tfan.tab}).
The setup of the simulations is the same as before, except that this time we take $256^3$ gridpoints over $[x,y,z]\in [\pm2.5,\pm2.5,\pm0.5]$.
This time we perturb the initial equilibrium (Eq.~(\ref{anis_null.eq})) by applying a rotational driving velocity on the boundaries around the spine footpoints both above and below the null, as in \cite{galsgaardpriest2003}. The sense of rotation is opposite above and below the fan plane. Specifically, we apply the following azimuthal velocity in the $z=\pm 0.5$ planes;
\EQ
v_\theta=A\left(\left(\frac{t-\tau}{\tau}\right)^4-1\right)^2 r(1+\tanh(1-C^2 r^2)) \qquad t\leq 2\tau
\EN
where $r=\sqrt{x^2+y^2}$, $C=10$, $\tau=1.6$ and $A=\mp 0.1$ at $z=\pm0.5$. We perturb the system using a different method to Sect.~\ref{tspine.num} for the following reasons.
First, the rotation of the fan plane (Sect.~\ref{tspine.num}) is done with an internal perturbation within the domain since driving from the boundary in this way in our Cartesian simulation domain would involve driving around the corners of the domain which is difficult to do in a smooth and stable way. Second, we choose to drive from the boundary in this section since applying an internal perturbation it is difficult to develop a significant current increase at the fan. This is because owing to the spatially uniform resistivity (which we cannot significantly reduce due to the limitations of  numerical resolution) an internal perturbation (which must be well localised initially) suffers significant diffusion before reaching the fan, as discussed by \cite{pontingalsgaard2007}. Note that this is not such an issue for the case where the fan field lines are rotated (Sect.~\ref{tspine.num}) since in that case the incoming pulse spreads only in one direction ($z$) owing to the hyperbolic nature of the field while it contracts in two ($x,y$) -- however, for the case of a spine rotation the in-coming pulse spreads in two directions ($x,y$) and contracts only in one ($z$).

\begin{figure} \centering
\centering
 \includegraphics[width=0.45\textwidth]{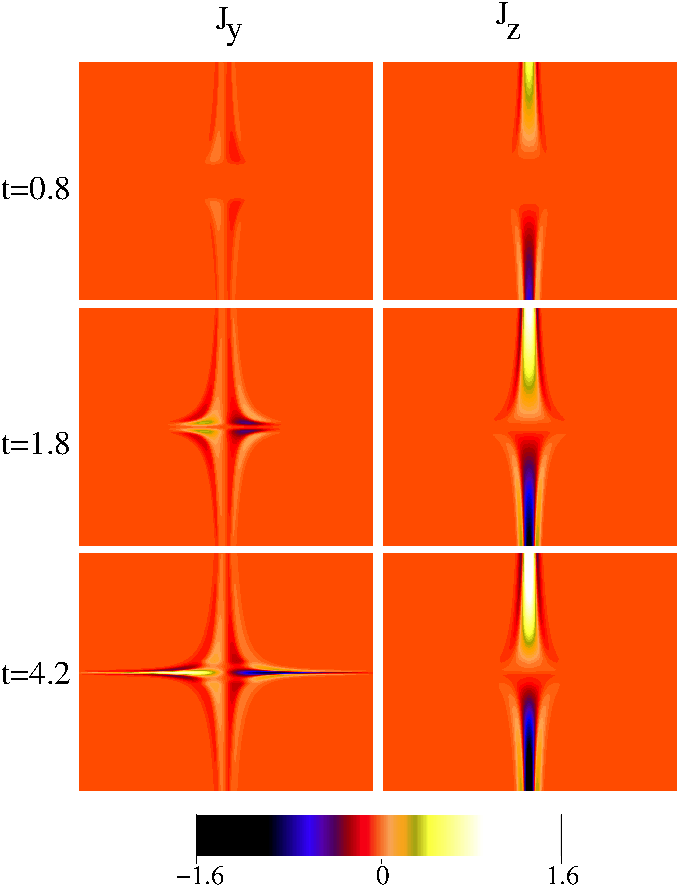}
 \caption[]{\label{jcomp_sprtd.fig}  Contours of $J_y$ (left) and $J_z$ (right) in the $x=0$ plane over $[y,z]\in[\pm2.5,\pm0.5]$ for $t=0.8$ (top), $t=1.8$ (middle) and $t=4.2$ (bottom), for the torsional fan simulation with $p=1$.  }
\end{figure}

For the symmetric case $p=1$, the evolution of the system
following the initiation of the driving velocity is described in detail by \cite{galsgaardpriest2003}. A disturbance that is dominated by a torsional Alfv{\' e}n wave is launched from the driving $z$ boundaries towards the null. The wave front spreads along the $x$- and $y$-directions travelling along the field lines, with its velocity in the $z$-direction being independent of $x$ and $y$. The wavefront steepens as it approaches the fan surface due to the hyperbolic structure of the field, eventually forming a planar current layer in the fan. This process is demonstrated in Fig.~\ref{jcomp_sprtd.fig} (see also the left-hand image in Fig.~\ref{jfan_sprtd.fig}), from which one can see that the current associated with the current layer is oriented parallel to the fan surface and flows radially inwards toward the null (note that in this $x=0$ plane, $J_x$ is zero by symmetry). It is worth noting that there also remains a large distributed current near the boundaries where the twisting was applied -- observed as strong concentrations of $J_z$ -- with the modulus of the current in this region being approximately equal to that of the current in the layer at the fan. We have repeated the simulation with different values of $\eta$, and found that for lower $\eta$ the peak current density at the fan is stronger relative to the concentration near the boundary. Hence, if we were able to run the simulation with a realistic value of $\eta$ for an astrophysical plasma the current layer at the fan would dominate. 

\begin{figure} \centering
\centering
 \includegraphics[width=0.45\textwidth]{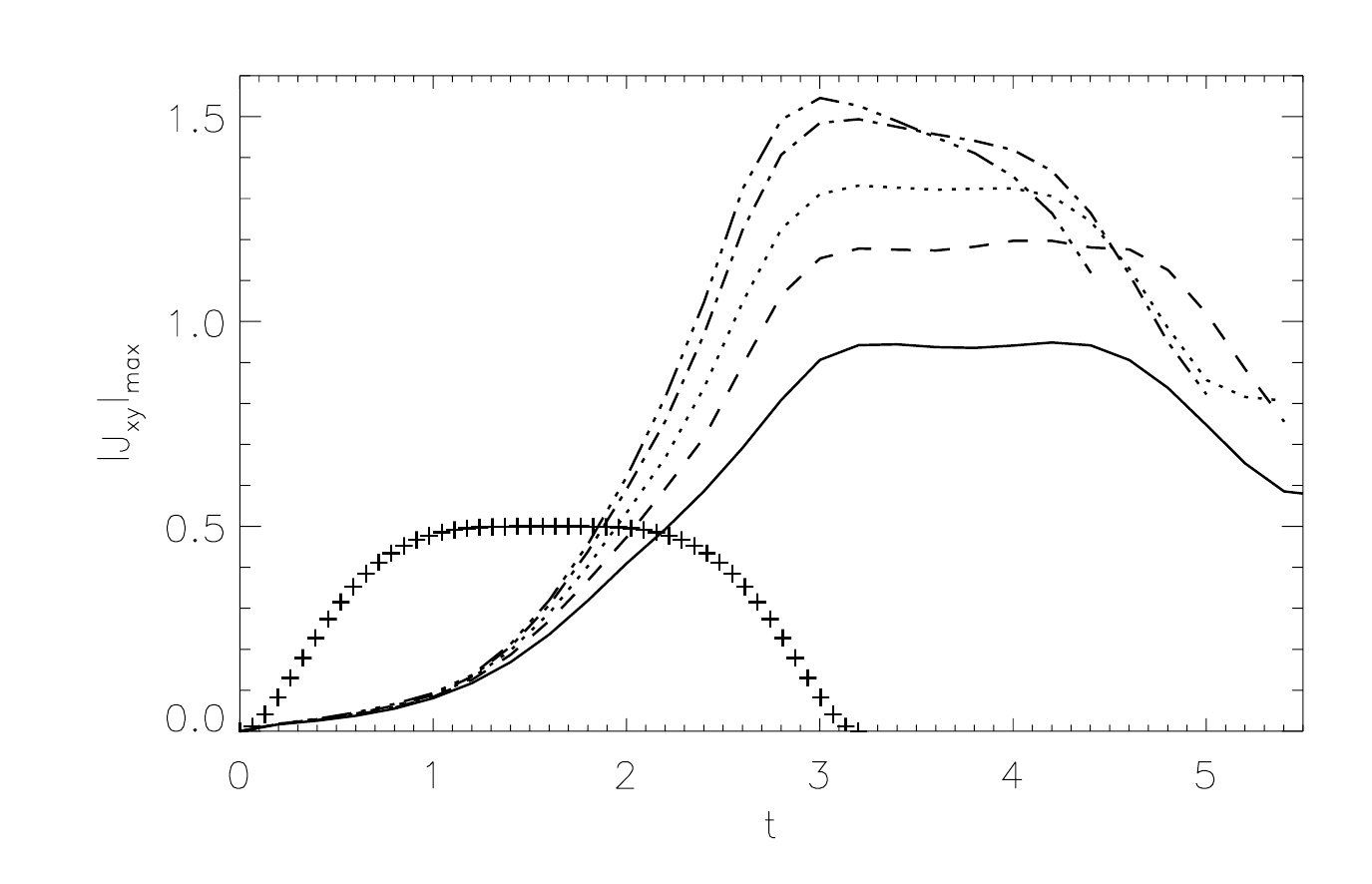}
\caption[]{\label{jmax_sprtd.fig}  Evolution of the maximum of $|\JJ_{xy}|$ for the different torsional fan simulation runs: $p=1$ (solid line), $p=2$ (dashed), $p=3$ (dotted), $p=5$ (dot-dashed) and $p=10$ (triple-dot-dashed). For comparison, the time variation of the amplitude of the driving velocity is also represented, with the crosses.}
\end{figure}

\begin{figure} \centering
\centering
 \includegraphics[width=0.7\textwidth]{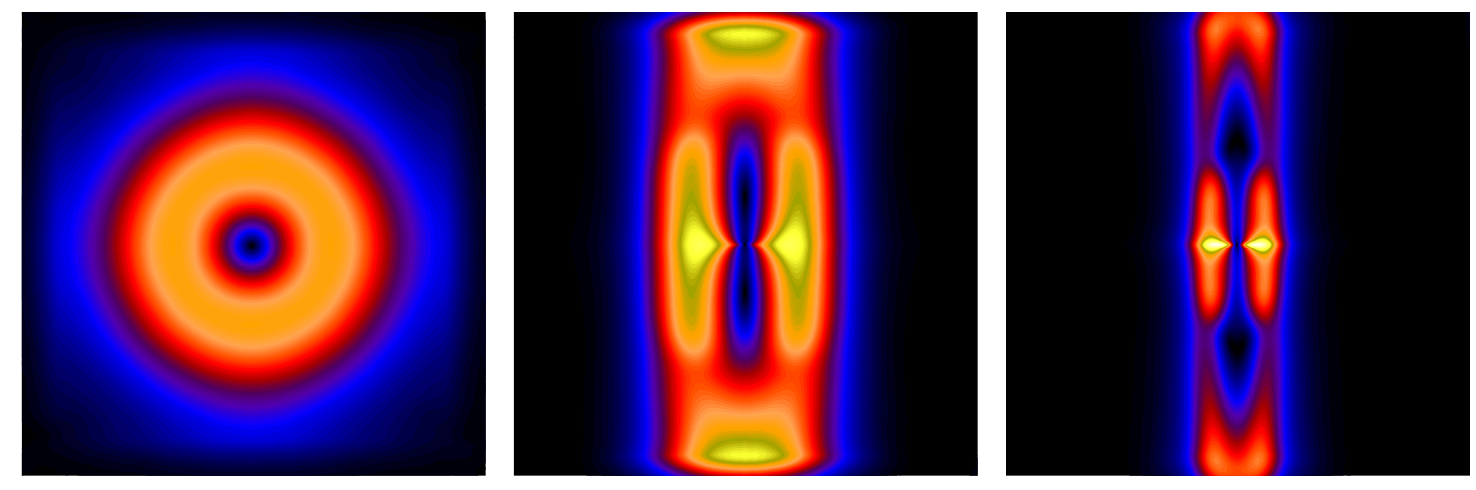}\\
 \includegraphics[width=0.3\textwidth]{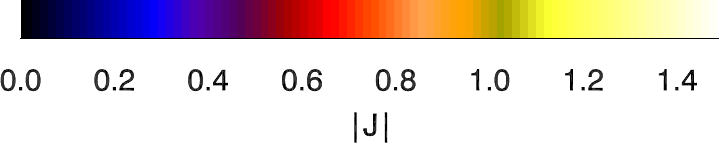}
\caption[]{\label{jfan_sprtd.fig}  Contours of $|{\bf J}|$ for the torsional fan simulations, plotted in the $z=0$ (fan) plane over $[x,y]\in[\pm2.5,\pm2.5]$ for $p=1$ (left), $p=2$ (middle) and $p=5$ (right). Taken in each case at the time when the value of $|{\bf J}_{xy}|$ reaches a maximum. }
\end{figure}

As a result of the dominance of the $J_z$ component near the driving boundaries (as discussed above), it is most clear to observe the formation and evolution of the fan current layer by plotting the evolution of the peak value of $|\JJ_{xy}|$, as in Fig.~\ref{jmax_sprtd.fig} (the dominant current component in the current layer is parallel to the fan surface for all $p$). We can see that for $p=1$ this quantity rises steadily as the pulse approaches the fan, after which there is a period of around $t=1.5$ when the value remains steady, after which it decreases. Note that this steady period is consistent with the period during which the driving velocity remains steady -- see the crosses in the figure. 
 It is clear that the overall qualitative behaviour is similar between the simulations. However, the overall peak current attained  is largest when the initial field is most asymmetric ($p=10$). Note that for the simulations with the largest values of $p$ the peak current does not remain steady for such a long period as for small $p$. This may be influenced by the fact that the disturbance interacts with the $y$-boundaries at later time for large $p$.

We now examine the spatial structure of the current distribution for simulation runs with different $p$. When the boundary driving is initiated the azimuthal symmetry of the wavefront that propagates into the domain is broken. The wavefront remains approximately planar ($z$ constant), spreading into an elliptical shape with long axis along the $y$-direction (along which the Alfv{\' e}n speed is greater) and short axis along $x$. However, as the wavefront gets closer to the null, it becomes steadily more inhomogeneous, with the current density focussing in the weak field regions near the $x$-axis. That is, although the current layer that forms eventually at the fan is more extended along $y$ (as a simple consequence of mapping a circular driving region on the boundary along $\BB$), this current is most intense along the short axis of the ellipse, as shown in Fig.~\ref{jfan_sprtd.fig}. As $p$ is increased, the peak current increases as described above, and this increasing current maximum is localised in a gradually narrower `channel' around the $x$-axis.

\begin{figure} \centering
\centering
 \includegraphics[width=0.45\textwidth]{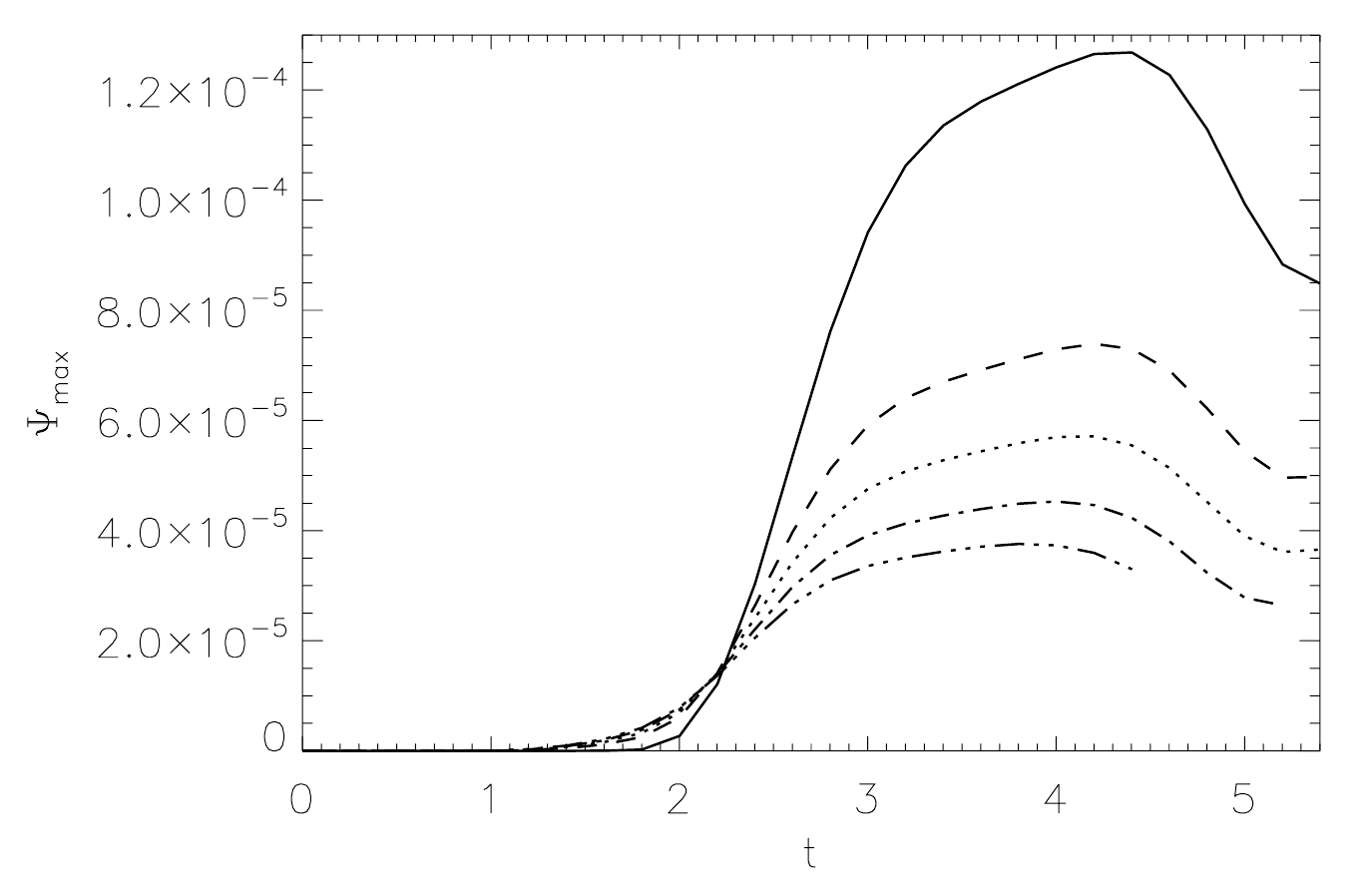}
\caption[]{\label{recr_sprtd.fig}  Evolution of the reconnection rate for the different torsional fan simulation runs: $p=1$ (solid line), $p=2$ (dashed), $p=3$ (dotted), $p=5$ (dot-dashed) and $p=10$ (triple-dot-dashed).}
\end{figure}

We turn now to consider the rate of reconnection in the different simulations. As discussed above, at the limited magnetic Reynolds number we are able to use the fan current layer does not dominate over the distributed current near the driving boundaries -- though indications are that it would for more realistic astrophysical parameters. Therefore, in order to examine only  the effect of reconnection in the thin current layer, we calculate the reconnection rate by integrating $E_\|$ along the magnetic field line in the fan plane that passes through the peak of the current density -- which in practice lies very close to the $x$-axis by symmetry. When the current layer has fully formed at the fan this genuinely measures the reconnection rate associated with the dynamically forming fan current layer. Clearly at earlier stages when the perturbation is yet to reach the fan there will still be a modest amount of reconnection, which is not measured by this approach. Therefore, when analysing the plots of reconnection rate versus time shown in Fig.~\ref{recr_sprtd.fig}, one should bear in mind that the curves do not accurately portray the reconnection rate prior to $t\approx 2.5$. We see that the peak reconnection rate associated with the fan current layer occurs in each of the simulations at $t\approx 4$, just before the peak current density starts its decline. A clear pattern emerges that the reconnection rate is greatest for the symmetric case with $p=1$, and steadily decreases for simulations with larger $p$. 

\subsection{Torsional fan reconnection: discussion}
The numerical simulations discussed above demonstrate the localisation of a rotational perturbation towards the fan of a non-symmetric linear 3D null, resulting in torsional fan magnetic reconnection. The planar current structure that forms around the fan qualitatively resembles that found in the kinematic steady-state model described in Sect.~\ref{tspine.kin}. In particular, the current is dominated by a component parallel to the fan where the current vector is directed (at its maximum intensity when $p\neq 1$) radially towards the null. Current isosurfaces have an elliptical shape, with the short axis of the ellipse aligned with the weak field direction in the fan plane, along which the current is most intense. The eccentricity of the ellipse increases as the magnetic field asymmetry increases. The plasma flow also has a similar qualitative structure in the kinematic model and simulations. This structure is that of a rotation along elliptical paths around the spine, with these elliptical paths closely following the current density contours. 

In the torsional fan reconnection simulations -- in contrast to the torsional spine case -- the peak reconnection rate depends strongly on the magnetic field asymmetry ($p$ parameter). This is because in the torsional fan case, the contribution to the integrand in the reconnection rate definition (\ref{recrate.eq}) comes largely from field lines closely aligned to the $xy$-plane, which is the plane in which the magnetic field varies as $p$ is varied. The interpretation of the reconnection rate when $p\neq 1$ should be the same as that described above in Sect.~\ref{tspine.discuss}.
Perhaps counter-intuitively, the scaling of the reconnection rate with $p$ is opposite to the dependence of the peak current on $p$ (as $p$ increases the peak current increases while the reconnection rate decreases -- see Figs.~\ref{jmax_sprtd.fig}, \ref{recr_sprtd.fig}). This is because, unlike in 2D, the reconnection rate is not a local quantity defined at a point, but rather is defined as an integral along a field line. So if the peak value of the integrand in Eq.~(\ref{recrate.eq}) is larger, but the integrand is non-zero over a shorter section of the field line, this may result in a lower net value for the integral, or {\it vice-versa}. Note that this behaviour was also observed in the study of spine-fan reconnection carried out by \cite{alhachami2010}. Note also that the effect was not captured in the kinematic solution -- however, it is worthwhile noting that the scaling of the reconnection rate with $p$ is much weaker than that of the peak current. Specifically, while the current increases by a factor $\approx 3$ between $p=1$ and $p=10$ in the kinematic solution, the reconnection rate increases only by a factor $\approx 1.2$.

In the simulations after the current layer that formed dynamically at the fan has started to dissipate, the distributed current near the driving boundaries dominates to a steadily greater extent, since the large-scale current concentrations dissipate only very slowly. This is consistent with the results of \cite{pontincraig2005}, who found that only when shear perturbations of the spine/fan are made does the lowest energy state of the system involve a current sheet at the null -- when rotational perturbations are performed around the spine then the lowest energy state is achieved when the twist of the field is distributed along the field lines, in concentrations that extend outwards from the driving boundaries.

Finally, note that here we have chosen to model an instance of torsional fan reconnection in which the vorticity of the driving flow around the spine has opposite sign on the opposite boundaries. As a consequence, a strong current -- directed predominantly in the radial direction -- is present at the fan surface.
However, in our kinematic model we could equally have chosen $B_\theta$ to be even in $z$ to model the situation where the driving flows have the same sign of vorticity, which leads to cancellation at the fan plane of the currents generated \citep{galsgaardpriest2003}. It should also be noted that the return currents present close to $z=\pm b$ are not present in the simulations as we drive only in one sense, but could be induced by reversing the sign of the driving velocity at some intermediate time in the simulations.

\section{Conclusion}
\label{conc.sec}
Here we have presented analytical and numerical models for torsional spine and torsional fan magnetic reconnection at 3D null points. The analytical models included for the first time fully localised current layers -- focussed at the spine or fan -- that determine the boundary of the non-ideal region, thus alleviating the requirement in previous models to have an artificially localised (`anomalous') resistivity. We also for the first time investigated the generic case where the null point is not radially symmetric, i.e.~where the fan eigenvalues are not equal.

3D null points have been demonstrated to be present in abundance in the solar corona, and the same is likely to be true in other astrophysical environments. Recent results from solar observations and from simulations suggest that reconnection at such 3D nulls may play an important role in the coronal dynamics. The fan separatrix surfaces of these 3D nulls divide the coronal magnetic field into distinct topological domains, i.e.~distinct regions of magnetic connectivity between the photospheric flux concentrations.
The torsional spine and torsional fan reconnection modes do not act to transfer magnetic flux between these topological domains, but rather permit a rotational slippage within the domains of the magnetic flux  lying close to the nulls and therefore the domain boundaries.
These null point reconnection modes are therefore unlikely to be involved in energetic events that involve a large-scale restructuring of the magnetic flux between topological domains as the  coronal field seeks a lower energy state (e.g.~during solar flares). Rather, they are a mechanism to dissipate energy and reduce stress associated with the dynamic perturbation of the coronal field by the turbulent boundary driving from the photosphere. 

We have shown that the geometry of the current layers within which torsional spine and torsional fan reconnection occur is strongly dependent on the symmetry of the magnetic field defining the null point. Torsional spine reconnection still occurs in a narrow tube around the spine, but with elliptical cross-section when the fan eigenvalues are different. The eccentricity of the ellipse increases as the degree of asymmetry increases, with the short axis of the ellipse being along the strong field direction. Furthermore, the current profile is not azimuthally symmetric around the spine, but is peaked in these strong field regions. The numerical simulations suggest that the spatiotemporal peak current, and the peak reconnection rate attained, do not depend strongly on the degree of asymmetry. For torsional fan reconnection, the reconnection occurs in a planar disk in the fan surface, which is again elliptical when the symmetry of the magnetic field is broken. The short axis of the ellipse is along the weak field direction, with the current being peaked in these weak field regions. The peak current and peak reconnection rate in this case {\it are} clearly dependent on the asymmetry, with the peak current increasing but the reconnection rate decreasing as the degree of asymmetry is increased.

While we have relaxed the rotational symmetry of the magnetic field in these studies, the field structure of a linear null is still relatively simple. In the future it will be important to understand how these reconnection modes are modified -- and how they release the energy associated with imposed stresses -- when the null point is embedded in a more realistic coronal geometry. One recent study suggests that other features present in the magnetic field may attract the current preferentially over the nulls and therefore may in some cases inhibit the formation of the torsional spine and torsional fan current layers \citep{santos2011}. {Other  studies have focussed on explaining explosive events and thus  have been motivated by  identifying reconnection via flux transfer between topological domains, facilitated by the spine-fan reconnection mode \citep[e.g.][]{masson2009,pariat2009,torok2009}. In practice one would expect a combination of the pure reconnection modes to appear, and significant further study is required to determine the manifestations of, interactions between, and relative importance of the spine-fan mode and the two torsional modes discussed here}. The importance of a number of other parameters such as the magnitude of the perturbation, the plasma-$\beta$ and the resistivity $\eta$ is also yet to be explored.

\section*{Acknowledgements}
The authors would like to acknowledge fruitful discussions with G.~Hornig, A.~Wilmot-Smith and A.~Yeates. D.~I.~P.~gratefully acknowledges support from the Royal Society. A.~K.~Al-H. was supported  by a grant from the Iraqi Government. Support by the European Commission through the Solaire Network (MTRNCT-2006-035484) is gratefully acknowledged. Computational simulations were run on the UKMHD Computing Consortium's Beowulf cluster.

\bibliographystyle{apalike} 

\end{document}